\documentclass{emulateapj}
\usepackage{graphicx}
\begin{document}

\def\etal{et al.\ \rm}
\newcommand{\ga}{\gtrsim}
\newcommand{\la}{\lesssim}

\title{Dynamical evolution of planetesimals in 
protoplanetary disks.}

\author{R. R. Rafikov}
\affil{IAS, Einstein Dr., Princeton, NJ 08540}
\email{rrr@ias.edu}

%%%%%%%%%%%%%%%%%%%%%%%%%%%%%%%%%%%%%%%%%%%%%%%%%%%%%%%%%%%

\begin{abstract}
The current picture of terrestrial planet formation 
relies heavily on our understanding of the dynamical
evolution of planetesimals --- asteroid-like bodies thought 
to be planetary building blocks. 
In this study we investigate 
the growth of eccentricities and inclinations of planetesimals
in spatially homogeneous protoplanetary disks using 
methods of kinetic theory. Emphasis is put on clarifying the 
effect of gravitational scattering between planetesimals 
on the evolution of their random velocities. 
We explore disks with a realistic mass spectrum of 
planetesimals evolving in time, similar to that 
obtained in self-consistent simulations 
of planetesimal coagulation: distribution scales 
as a power law of mass for small planetesimals and is
supplemented by an extended tail of bodies at large masses 
representing the ongoing runaway growth in the system.
We calculate the behavior of planetesimal random velocities 
as a function of the planetesimal mass spectrum both analytically 
and numerically; results obtained by the two approaches agree 
quite well. Scaling of random velocity 
with mass can always be represented as a combination of power laws 
corresponding to different velocity regimes (shear- or 
dispersion-dominated) of planetesimal gravitational 
interactions. For different mass spectra we calculate analytically
the exponents and time dependent normalizations of these power 
laws, as well as the positions of the transition regions between 
different regimes. 
It is shown that random energy equipartition between 
different planetesimals can only be achieved in 
disks with very steep mass distributions (differential surface 
number density of planetesimals falling off steeper than 
$m^{-4}$), or in the runaway tails. In systems with 
shallow mass spectra (shallower than $m^{-3}$) 
random velocities of small planetesimals 
turn out to be independent of their masses.
We also discuss the damping effects of inelastic collisions 
between planetesimals and of gas drag, and their importance
in modifying planetesimal random velocities. 
\end{abstract}


%%%%%%%%%%%%%%%%%%%%%%%%%%%%%%%%%%%%%%%%%%%%%%%%%%%%%%%%%%%
%%%%%%%%%%%%%%%%%%%%%%%%%%%%%%%%%%%%%%%%%%%%%%%%%%%%%%%%%%%

\section{Introduction.\label{sect:intro}}

Gravity-assisted agglomeration of planetesimals in the protoplanetary
disks around young stars is a backbone of a currently widely accepted 
paradigm of terrestrial planet formation, despite many uncertainties 
in the details of specific processes involved. In many respects our 
understanding of the planetesimal disk evolution heavily relies on the  
numerical investigations, and these studies significantly 
grew in complexity 
in recent years (Wetherill \& Stewart 1989, 1993; Kenyon \& Luu 1998, 
2002; Inaba \etal 2001). On one hand, the sheer scale of 
simulated systems has expanded considerably, which was made possible by
the increase in computing power; on the other hand, a wider range of 
concomitant physical phenomena (fragmentation, migration, etc.) 
can now be routinely followed simultaneously with the evolution of 
planetesimal mass and velocity distributions. 

Unfortunately, very often this complexity creates a problem
when one tries to understand a {\it relative} role of each of 
the specific processes in shaping the planetesimal mass and velocity 
spectra. Disentangling the contributions of different physical
nature in the results of numerical simulations 
certainly is a tantalizing task which can often yield very confusing
conclusions. Thus it is very important to be able to isolate different 
physical processes operating in the planetesimal disks 
and to study their effects on the disk evolution separately. 
One way to do this is to build simple but realistic models which
are easy to analyze and at the same time are able to grasp 
the main features of the physics involved. Such approach would 
grant us good analytical understanding of relevant processes,
and we are going to follow this route in our study.

Behavior of planetesimal velocities is one of the
crucial ingredients of the terrestrial planet formation picture because of 
the very important role played by the gravitational focusing (which 
depends on planetesimal velocities) in assembly of the big bodies 
in planetesimal disks. The purpose of this paper is to explore the 
evolution of inclinations and eccentricities (which represent 
planetesimal random velocities) caused by the mutual 
gravitational scattering of planetesimals. This is a clean and 
well-posed problem. The effects of gas 
drag and inelastic collisions are initially neglected 
but later on we discuss how their inclusion affects our results.
To keep things as clear as possible
we have opted not to try coupling self-consistently 
planetesimal velocity evolution to the evolution of 
the mass spectrum in this study.
Instead, we follow changes of random velocities
in a disk with a distribution of planetesimal sizes 
which evolves in some {\it prescribed} manner. 
However simplistic this assumption seems to be, 
in adopting it we try to use mass spectra which are similar to 
those produced by self-consistent 
coagulation simulations, and to study a wide variety of such mass 
distribution models. 

Another argument in favor of this approach is the 
fact that the dynamical relaxation time in a disk of gravitationally 
interacting planetesimals is much shorter than the physical collision
timescale (characterizing evolution of the mass spectrum) if 
relative random velocities of planetesimals are smaller than the 
escape velocities from their surfaces (which is very often the case). 
The dynamical evolution of planetesimal disk can then be 
approximated as that of a system with a quasi-stationary planetesimal 
mass spectrum. Understanding this problem is a logical  
step necessary to provide a clearer perspective on how 
to build a fully self-consistent theory of planet formation. 

We describe the statistical approach to the problem of the 
evolution of planetesimal random velocities in \S \ref{sect:velev}.
Our model of planetesimal mass spectrum is introduced \S 
\ref{sect:mass_spec}. In \S \ref{sect:scaling} we outline the results for
the distribution of planetesimal random velocities vs. planetesimal
 masses as functions of the input mass distribution and time; 
we also compare these analytical predictions with numerical results. 
Table \ref{table2} summarizes our major findings in a concise form.
In \S \ref{sect:dissipative} we comment on how gas drag and inelastic 
collisions between planetesimals can affect planetesimal velocity 
spectra. We conclude in \S \ref{sect:disc} by the discussion of our 
results and comparison with previous studies.

%%%%%%%%%%%%%%%%%%%%%%%%%%%%%%%%%%%%%%%%%%%%%%%%%%%%%%%%%%%
%%%%%%%%%%%%%%%%%%%%%%%%%%%%%%%%%%%%%%%%%%%%%%%%%%%%%%%%%%%

\section{Velocity evolution equations.
\label{sect:velev}}

When the number of planetesimals in the protoplanetary disk
is very large and their masses are not 
sufficient to induce strong local nonuniformities in the disk 
(Ida \& Makino 1993; Rafikov 2001, 2003b, 2003c), 
statistical approach and homogeneous ``particle in a box'' assumption 
are very helpful in the treatment of planetesimal evolution
(Wetherill \& Stewart 1989). We assume that 
planetesimals with mass $m$ have velocity dispersions of eccentricity and 
inclination $\sigma_e(m,t)$ and $\sigma_i(m,t)$ associated with them. 
It is natural to characterize planetesimal disk by its mass distribution, and 
we assume that $N(m,t)dm$ is the {\it surface number density} of planetesimals 
with masses between $m$ and $m+dm$.

In studying the dynamics of planetesimal disks we make the following
natural assumptions:
\begin{itemize}

\item disk is locally homogeneous in radial direction and azimuthally 
symmetric; it is in Keplerian rotation around central star,

\item epicyclic excursions of planetesimals in horizontal and vertical
directions are small compared with the distance to the 
central object,

\item planetesimal masses are much smaller than the mass of the 
central object $M_c$.
\end{itemize}

Two important consequences immediately follow from this set of assumptions.
First, the gravitational scattering of planetesimals during their 
encounter can be studied in the framework of Hill approximation
(H\'enon \& Petit 1986; Ida 1990; Rafikov 2003a). In this 
approach, gravitational 
interaction between two bodies with masses $m$ and $m^\star$ introduces a 
natural lengthscale --- Hill radius\footnote{This definition 
follows original work of H\'enon \& Petit (1986) and 
differs from $R_H\equiv a[(m+m^\star)/3M_c]^{1/3}$ often used in the 
literature (e.g. Ida 1990; Stewart and Ida 2000).} 
\begin{eqnarray}
R_H\equiv a\left(\frac{m+m^\star}{M_c}\right)^{1/3}\nonumber\\
=10^{-4}~{\rm AU}~
a_{AU}\left(\frac{m+m^\star}{2\times 10^{21}~g}\right)^{1/3},
\label{eq:hill}
\end{eqnarray}
where $a$ is the distance from the central object.
Numerical estimate made in (\ref{eq:hill}) assumes $M_c=M_\odot$ and
$a_{AU}\equiv a/(1~\rm{AU})$, and is intended to illustrate that
typically $R_H\ll a$.
Hill approximation yields two significant simplifications:
\begin{itemize}

\item The outcome of the interactions between two bodies depends only on
their {\it relative} velocities and distances.

\item If all relative distances are scaled by $R_H$ and relative velocities
of interacting bodies are scaled by $\Omega R_H$, then the outcome of the 
gravitational scattering depends only on the initial values of 
scaled relative quantities.
\end{itemize}

Second, numerical studies (Greenzweig \& Lissauer 1992; Ida \& Makino 1992) 
have demonstrated that in homogeneous 
planetesimal disks the distribution function $\psi(e,i)$
of absolute values of planetesimal 
eccentricities $e$ and inclinations $i$  
is well represented by the  Rayleigh distribution:
\begin{equation}
\psi(e,i)de di=\frac{e de 
~i di}{\sigma_{e}^2
\sigma_{i}^2}\exp\left[-\frac{e^2}{2\sigma_{e}^2}
-\frac{i^2}{2\sigma_{i}^2}\right],
\label{eq:Gauss}
\end{equation}
where $\sigma_e$ and $\sigma_i$ are the aforementioned dispersions of 
eccentricity and inclination. It also follows from azimuthal 
symmetry that horizontal and vertical 
epicyclic phases $\tau$ and $\omega$ are distributed uniformly 
in the interval $(0,2\pi)$. These facts have important 
ramifications as we demonstrate below.

Let's consider the gravitational interaction of two planetesimal 
populations: one with mass $m$ and eccentricity and inclination 
dispersions $\sigma_e$ and $\sigma_i$ and the other with mass $m^\star$ 
and dispersions $\sigma_e^\star$ and $\sigma_i^\star$. 
When the velocity\footnote{Throughout the paper we often refer to 
eccentricities and inclinations of planetesimals as their random 
velocities.} distribution function of planetesimals has the form 
(\ref{eq:Gauss}) it can be shown (Rafikov 2003a) 
that the evolution of $\sigma_e$ and 
$\sigma_i$ due to the gravitational interaction with population of 
mass $m^\star$ proceeds according to 
\begin{eqnarray}
&& \frac{\partial \sigma_e^2}{\partial t}=
|A|N^\star a^2\left(\frac{m+m^\star}{M_c}\right)^{4/3}
\frac{m^\star}{m+m^\star}\nonumber\\
&& \times\left[
\frac{m^\star}{m+m^\star}H_1
+2\frac{\sigma_e^2}{\sigma_e^2+ 
\sigma_e^{\star 2}}H_2\right],\nonumber\\
&& \frac{\partial \sigma_i^2}{\partial t}=
|A|N^\star a^2\left(\frac{m+m^\star}{M_c}\right)^{4/3}
\frac{m^\star}{m+m^\star}\nonumber\\
&& \times\left[
\frac{m^\star}{m+m^\star}K_1
+2\frac{\sigma_i^2}{\sigma_i^2+ 
\sigma_i^{\star 2}}K_2\right],
\label{eq:homog_heating}
\end{eqnarray}
where $N^\star$ is the surface number density of planetesimals of mass 
$m^\star$,   
and $A=-(r/2)d\Omega/dr$ is a measure of shear in the disk
[in Keplerian disks $A=-(3/4)\Omega$]. 
Scattering coefficients $H_{1,2}$ are defined as
\begin{eqnarray}
&& H_1=H_1\left(\tilde \sigma_{e r},
\tilde \sigma_{i r}\right)\nonumber\\
&& =
\int d{\bf \tilde e}_r d{\bf \tilde i}_r \tilde
\psi_r({\bf \tilde e}_r,{\bf \tilde i}_r)
\int\limits_{-\infty}^{\infty}d\tilde h 
|\tilde h|\left(\Delta {\bf \tilde e}_{sc}\right)^2,
\nonumber\\
&& H_2=H_2\left(\tilde \sigma_{e r},
\tilde \sigma_{i r}\right)\nonumber\\
&& =\int d{\bf \tilde e}_r d{\bf \tilde i}_r \tilde
\psi_r({\bf \tilde e}_r,{\bf \tilde i}_r)
\int\limits_{-\infty}^{\infty}d\tilde h |\tilde h|
({\bf \tilde e}_r\cdot\Delta {\bf \tilde e}_{sc}),
\label{eq:stirring_coeffs}
\end{eqnarray}
where $\tilde \sigma_{er,ir}=(\sigma_{e,i}^2+\sigma_{e,i}^{\star 2})^{1/2} 
a/R_H$ are the dispersions of relative eccentricity and inclination 
normalized in Hill coordinates [see (\ref{eq:hill})], $\tilde h$ is a semimajor axes 
difference of interacting bodies scaled by $R_H$, and $\tilde {\bf e}_r,
\tilde {\bf i}_r$ are the relative eccentricity and inclination {\it vectors} in 
Hill units (Goldreich \& Tremaine 1980; Ida 1990). 
Function $\Delta {\bf \tilde e}_{sc}$ represents a change of 
$\tilde {\bf e}_r$ produced in the course of scattering,
which is a function of not only absolute values of $\tilde e_r, \tilde i_r$,
and $\tilde h$, but also of the relative epicyclic phases
(for which reason we use vector eccentricities and inclinations here). 
The distribution function of relative eccentricities 
and inclinations of planetesimals 
$\tilde \psi_r({\bf \tilde e}_r,{\bf \tilde i}_r)$ is analogous in its functional 
form to (\ref{eq:Gauss}), but with $\sigma_{e,i}$ replaced by  
{\it relative} velocity dispersions $\tilde \sigma_{er,ir}$. 
Expressions analogous to (\ref{eq:stirring_coeffs})
can be written down also for inclination scattering coefficients $K_1$ and
$K_2$. Note that the only assumption used in deriving 
(\ref{eq:homog_heating}) is that of the specific form   (\ref{eq:Gauss})
of the distribution function of planetesimal
eccentricities and inclinations (Rafikov 2003a).

Equations (\ref{eq:homog_heating}) describe dynamical evolution 
driven only by gravitational scattering of planetesimals, implicitly assuming
that they are point masses. Physical size of planetesimal 
$r_p\approx 3\times 10^{-7}~{\rm AU}~(m/(10^{21}~\rm{g}))^{1/3}$ (for physical density
$3~\rm{g}~\rm{cm}^{-3}$)
is very small compared to the corresponding Hill radius
(\ref{eq:hill}), which often justifies the neglect of physical 
collisions between planetesimals. However, at high relative
velocities --- higher than the escape speed from planetesimal surfaces ---
one can no longer disregard highly inelastic physical collisions which 
strongly damp planetesimal random motions. In this study
we proceed by assuming first that velocities of planetesimals are below their 
escape velocities, and then discussing in \S \ref{sect:dissipative} 
how abandoning this assumption affects our conclusions.

Equation (\ref{eq:homog_heating}) and definitions (\ref{eq:stirring_coeffs})
have been previously derived by other authors in a slightly different 
form  
(Wetherill \& Stewart 1988; Ida 1990; Tanaka \& Ida 1996; Stewart \& Ida 2000): 
\begin{eqnarray}
&& \frac{\partial \sigma_e^2}{\partial t}=
|A|N^\star a^2\left(\frac{m+m^\star}{M_c}\right)^{4/3}
\frac{m^\star}{m+m^\star}\nonumber\\
&& \times\left[
\frac{m^\star}{m+m^\star}(H_1+2H_2)
+2\frac{m}{m+m^\star}\frac{\sigma_e^2}{\sigma_e^2+ 
\sigma_e^{\star 2}}H_2\right.\nonumber\\
&& \left.
-2\frac{m^\star}{m+m^\star}\frac{\sigma_e^{\star 2}}
{\sigma_e^2+\sigma_e^{\star 2}}H_2
\right].
\label{eq:Ida_form}
\end{eqnarray} 
In this form the first term in brackets on the r.h.s. describes the so-called
{\it viscous stirring} which is proportional to the phase space average of 
$\Delta ({\bf e}_r^2)$ (see definitions of $H_1$ and $H_2$). 
Depending on $\tilde \sigma_{e r}$ and $\tilde \sigma_{i r}$ it can be 
either positive or negative. Second term represents 
the phenomenon of {\it dynamical friction} well known from galactic dynamics. 
In the limit $m\gg m^\star$ its contribution is proportional to the mass of particle 
under consideration and to the surface mass density of field particles, but is 
independent of individual masses of field particles (cf. Binney \& Tremaine 1987).
This term is negative since it represents the gravitational interaction 
of a moving body with the wake of field particles formed {\it behind} it as a 
result of gravitational focusing; thus $H_2<0$ and $K_2<0$. Finally, the third 
term describes the increase of random velocities of a particular body 
 at the expense of random motion of field planetesimals it interacts with.
This effect is analogous to the first order Fermi mechanism of the 
cosmic ray acceleration via scattering of energetic particles by randomly moving 
magnetic field inhomogeneities (Fermi 1949). 
Note that in previous studies of planetesimal scattering 
it is the combination of the second and third terms on the r.h.s. of 
(\ref{eq:Ida_form}) that is called the dynamical friction
(Stewart \& Wetherill 1988; Ida 1990). The combined effect of these two
terms is to drive planetesimal system to equipartition of random energy 
between planetesimals of different mass (which would be realized in the
absence of viscous stirring).

In this work we analyze planetesimal velocity evolution with the aid of 
equation (\ref{eq:homog_heating}). We call the first
(positive) term on its r.h.s. {\it gravitational 
stirring} or {\it heating} (different from viscous stirring), 
while second (negative) term is called {\it gravitational friction} or
{\it cooling} (different from dynamical friction\footnote{I am grateful 
to Peter Goldreich and Re'em Sari for pointing this out to me.}). 
In this sense terms ``stirring'' and ``friction'' are only used to describe 
positive and negative contributions to the growth rate of 
planetesimals random motion. Use of equations (\ref{eq:homog_heating})
rather than (\ref{eq:Ida_form}) has two
obvious advantages: (1) gravitational stirring and friction  
have different dependences on $m^\star/(m+m^\star)$ which considerably 
simplifies analysis of velocity evolution equations, and (2)
stirring and friction terms have definite signs (unlike viscous stirring 
and dynamical friction used in previous studies of planetesimal dynamics).

Gravitational scattering of planetesimals can proceed in two 
rather different velocity regimes, 
shear-dominated and dispersion-dominated. The former one 
is realized when $\sigma_{e,i}^2+\sigma_{e,i}^{\star 2}\ll (R_H/a)^2$, while
the latter holds when $\sigma_{e,i}^2+\sigma_{e,i}^{\star 2}\gg (R_H/a)^2$.
Analytical arguments and numerical calculations demonstrate that in the 
dispersion-dominated regime $\sigma_e$ and $\sigma_i$ are of the same order and 
evolve in a similar fashion (e.g. Ida 1990; Stewart \& Ida 2000). 
This happens because scattering in this regime has a 
three-dimensional character making different components of velocity 
ellipsoid comparable to each other.
Thus, rough idea of the dynamical evolution in this case can be 
obtained by studying only one of the equations (\ref{eq:homog_heating}). 
Moreover, the behavior of the 
scattering coefficients in the dispersion-dominated 
regime can be calculated analytically 
 as a function of $\tilde \sigma_{e r}$
and $\tilde \sigma_{i r}$ in two-body approximation 
(Stewart \& Wetherill 1988; Tanaka \& Ida 1996), and one finds that
\begin{eqnarray}
&& \left\{
\begin{array}{l}
H_1\\
K_1
\end{array}
\right\}
=
\left\{
\begin{array}{l}
A_1\left(\tilde \sigma_{i r}/\tilde \sigma_{e r}\right)\\
C_1\left(\tilde \sigma_{i r}/\tilde \sigma_{e r}\right)
\end{array}
\right\}
\frac{\ln\Lambda}{\tilde \sigma_{e r}^2},\nonumber\\
&& 
\left\{
\begin{array}{l}
H_2\\
K_2
\end{array}
\right\}
=-
\left\{
\begin{array}{l}
A_2\left(\tilde \sigma_{i r}/\tilde \sigma_{e r}\right)\\
C_2\left(\tilde \sigma_{i r}/\tilde \sigma_{e r}\right)
\end{array}
\right\}
\frac{\ln\Lambda}{\tilde \sigma_{e r}^2},
\label{eq:h1h2}
\end{eqnarray}
where $A_{1,2}, C_{1,2}$ are positive functions of 
inclination to eccentricity ratio 
$\tilde \sigma_{i r}/\tilde \sigma_{e r}$, 
and $\ln\Lambda$ is a Coulomb logarithm (Binney \& Tremaine 1987),
which in our case can be represented as (Rafikov 2003a)
\begin{eqnarray}
\Lambda\approx a_1\tilde\sigma_{i r}(\tilde \sigma_{e r}^2+a_2\tilde \sigma_{i r}^2),
\label{eq:Coulomb}
\end{eqnarray}
with $a_{1,2}$ being some constants. Explicit analytical expressions for coefficients
$A_{1,2}, C_{1,2}$ 
 have been derived by Stewart \& Ida (2000). 
One can deduce a specific useful property of these functions by 
considering a single-mass planetesimal population: in dispersion-dominated regime
the distribution of random energy between the vertical and horizontal motions
tends to reach a quasi-equilibrium state (see e.g. Ida \& Makino 1992).
Then the ratio $\sigma_{i r}/\sigma_{e r}$ 
is almost constant and both $\sigma_{e r}$ and   
$\sigma_{i r}$ grow with time, meaning that [see (\ref{eq:homog_heating})
\& (\ref{eq:h1h2})]
\begin{eqnarray}
A_1>2A_2,~~~~~~C_1>2C_2.
\label{eq:property}
\end{eqnarray}
This property will be used later in \S \ref{sect:scaling} 
\& \ref{app:subsect:steep}.

In the shear-dominated regime $\sigma_e$ and $\sigma_i$ 
evolve along different routes: eccentricity is excited much 
stronger than inclination because in this regime disk is 
geometrically thin and forcing in the plane of the disk is stronger
than in perpendicular direction. Eccentricity 
evolution is also quite rapid in this case 
because of the vigorous scattering (relative horizontal velocity increases by 
$\sim \Omega R_H$ in each synodic passage of two bodies initially 
separated by $\sim R_H$ in 
semimajor axes). Simple reasoning confirmed by 
numerical experiments suggests the following behavior of 
scattering coefficients
in the shear-dominated regime:
\begin{eqnarray} 
&& H_1(\tilde \sigma_{e r},\tilde \sigma_{i r})\approx B_1,~~~
H_2(\tilde \sigma_{e r},\tilde \sigma_{i r})\approx
-B_2 \tilde \sigma_{e r}^2,\nonumber\\
&& K_1(\tilde \sigma_{e r},\tilde \sigma_{i r})\approx D_1\tilde \sigma_{i r}^2,
~~~K_2(\tilde \sigma_{e r},\tilde \sigma_{i r})\approx 
-D_2 \tilde \sigma_{i r}^2,
\label{eq:cold_coeffs}
\end{eqnarray}
where $B_{1,2}, D_{1,2}$ are some positive constants, which 
can be fixed using numerical orbit 
integrations, see Appendix \ref{app:numerical_procedure}.
Simple qualitative derivation of (\ref{eq:h1h2}) and 
(\ref{eq:cold_coeffs}) can be found in Ida \& Makino (1993) 
and Rafikov (2003c).

For further convenience, we now reduce evolution equations to 
a dimensionless form. We relate planetesimal 
masses to the smallest planetesimal mass\footnote{Protoplanetary 
disks should contain planetesimals of different sizes but we assume 
that bodies lighter than $m_0$ are not 
important for the disk evolution. In the case considered here 
the choice of $m_0$ is dictated by the 
mass at which distribution would depart from the given power 
law form towards shallower scaling 
with mass (see also \S 4.1.1).} $m_0$ 
by using dimensionless quantity $x=m/m_0$. Instead of $N(m)$ we
introduce dimensionless mass distribution function $f(x)$ such that
\begin{eqnarray}
N(m)=\frac{\Sigma_p}{m_0^2}f(x)~~\mbox{and}~~
N(m)dm=\frac{\Sigma_p}{m_0}f(x)dx,
\label{eq:f_def}
\end{eqnarray}
where $\Sigma_p$ is the total {\it mass} surface density of planetesimals 
in the disk, which is a conserved quantity. 
We also rescale eccentricity and inclination dispersions: 
\begin{eqnarray}
s=\sigma_e\left(\frac{m_0}{M_c}\right)^{-1/3},~~~~
s_z=\sigma_i\left(\frac{m_0}{M_c}\right)^{-1/3}.
\label{eq:s_def}
\end{eqnarray}
This is equivalent to rescaling all distances by Hill radius
of smallest planetesimals.
In this notation the boundary between the shear- and 
dispersion-dominated regimes is given by conditions
\begin{eqnarray}
s^2+s^{\star 2}\approx (x+x^\star)^{2/3},~~
s_z^2+s_z^{\star 2}\approx (x+x^\star)^{2/3}.
\end{eqnarray}
We also introduce dimensionless time $\tau$ by
\begin{eqnarray}
\tau=\frac{t}{t_0},~~~\mbox{where}~~~t_0=|A|^{-1}\frac{m_0}{\Sigma_p a^2}
\left(\frac{m_0}{M_c}\right)^{-2/3}
\label{eq:tau}
\end{eqnarray}
is a timescale for the stirring of
planetesimal disk composed of bodies of mass $m_0$ only, 
on the boundary between the shear- and dispersion-dominated regimes.
Numerically, for $M_c=M_\odot$ one finds that
\begin{eqnarray}
t_0\approx 16~{\rm yr}~a_{AU}\left(\frac{10~\rm{g~cm}^{-2}}
{\Sigma_{p 0}}\right)\left(\frac{m_0}{10^{21}\rm{g}}\right)^{1/3},
\label{eq:time_estimate}
\end{eqnarray}
where we assumed that $\Sigma_p(a)=\Sigma_{p 0}a_{AU}^{-3/2}$
(Hayashi 1981). Dynamical timescale is rather 
short when $s,s_z\sim 1$, but it grows very rapidly as epicyclic 
velocities of planetesimals increase ($t\sim t_0 s^4$ in the 
dispersion-dominated regime).  

Random velocities of planetesimals of mass $x$ evolve due to the 
interaction
with all other bodies spanning the whole mass spectrum. 
Using our dimensionless notation we can rewrite equations 
(\ref{eq:homog_heating}), (\ref{eq:h1h2}), \& 
(\ref{eq:cold_coeffs}) in the following general 
form:
\begin{eqnarray}
\frac{\partial s^2}{\partial \tau}=
\int\limits_1^{\infty}dx^\star f(x^\star)
x^\star(x+x^\star)^{1/3}\nonumber\\
\times\left[\frac{x^\star}{x+x^\star} H_1
+2\frac{s^2}{s^2+ 
s^{\star 2}}H_2\right],
\label{eq:homog_heating_integr0.5}\\
\frac{\partial s_z^2}{\partial \tau}=
\int\limits_1^{\infty}dx^\star f(x^\star)
x^\star(x+x^\star)^{1/3}\nonumber\\
\times\left[\frac{x^\star}{x+x^\star} K_1
+2\frac{s_z^2}{s_z^2+ 
s_z^{\star 2}}K_2\right],
\label{eq:homog_heating_integr}
\end{eqnarray}
where $s\equiv s(x), s^\star\equiv
s(x^\star)$ and similarly for $s_z,s_z^\star$; 
in the shear-dominated regime 
[$s^2+s^{\star 2}, s_z^2+s_z^{\star 2}\ll (x+x^\star)^{2/3}$]
\begin{eqnarray}
&& H_1\approx C_1,~H_2\approx -C_2\frac{s^2+s^{\star 2}}
{(x+x^\star)^{2/3}},\nonumber\\
&& K_1\approx D_1\frac{s_z^2+s_z^{\star 2}}
{(x+x^\star)^{2/3}},~K_2\approx -D_2\frac{s_z^2+s_z^{\star 2}}
{(x+x^\star)^{2/3}},
\label{eq:shear}
\end{eqnarray}
and in the dispersion-dominated regime  
[$s^2+s^{\star 2}, s_z^2+s_z^{\star 2}\gg (x+x^\star)^{2/3}$]
\begin{eqnarray}
&& \left(
\begin{array}{l}
H_1\\
K_1
\end{array}
\right)
= \left(
\begin{array}{l}
A_1\\
B_1
\end{array}
\right)\frac{{(x+x^\star)^{2/3}}}
{s^2+s^{\star 2}}\ln\Lambda,\nonumber\\
&& \left(
\begin{array}{l}
H_2\\
K_2
\end{array}
\right)
= -\left(
\begin{array}{l}
A_2\\
B_2
\end{array}
\right)\frac{{(x+x^\star)^{2/3}}}
{s^2+s^{\star 2}}\ln\Lambda.
\label{eq:d-d}
\end{eqnarray}

We explore this system 
in \S \ref{sect:scaling} using two approaches: asymptotic analysis 
utilizing analytical methods, and direct numerical calculation of
velocity evolution. 

%%%%%%%%%%%%%%%%%%%%%%%%%%%%%%%%%%%%%%%%%%%%%%%%%%%%%%%%%%
%%%%%%%%%%%%%%%%%%%%%%%%%%%%%%%%%%%%%%%%%%%%%%%%%%%%%%%%%%

\section{Mass spectrum.
\label{sect:mass_spec}}

Starting from (\ref{eq:homog_heating_integr0.5})-(\ref{eq:d-d})
one would like to obtain 
the behavior of $s$ and $s_z$ as functions
of $x$ and $\tau$, given some planetesimal mass spectrum $f(x)$
as an input. In general,  
to do this one has to evolve equations
(\ref{eq:homog_heating_integr})-(\ref{eq:d-d}) numerically. However, it is 
usually true that the planetesimal size distribution 
spans many orders of magnitude 
in mass; thus one would expect that some general predictions can 
be made analytically about the {\it asymptotic} properties of planetesimal 
velocity spectrum. 

In this study we assume that planetesimal mass
distribution has a ``self-similar'' form; specifically we take
\begin{eqnarray}
f(x,\tau)=\psi(\tau)\varphi\left(\frac{x}{x_c(\tau)}\right),
\label{eq:s_sim}
\end{eqnarray} 
where $x_c(\tau)\gg 1$ is some fiducial planetesimal mass  
which steadily grows in time as a result of coagulation, 
$\psi(\tau)$ is a temporal modulation, and
function $\varphi$ represents a self-similar shape of the mass 
spectrum. We also assume in this study that asymptotically 
\begin{eqnarray}
\varphi(y)\sim \left\{
\begin{array}{l}
y^{-\alpha},~~~~~~~~~~~~~~~~~~~y\ll 1,\\
\exp(-y),~~~~~~~~~~~~~y\gg 1,
\end{array}
\right.
\label{eq:mass_spec}
\end{eqnarray}
where $\alpha>0$.
Such scaling behavior is often found in coagulation simulations.
We will see in \S \ref{sect:scaling} that for clarifying the 
asymptotic properties 
of planetesimal velocity spectrum it is enough to know only the asymptotic 
behavior of $\varphi(y)$ and not its exact shape. Moreover, in 
\S \ref{sect:disc} we demonstrate how our results for power-law size 
distributions can be generalized for other mass spectra.

Normalization $\psi(\tau)$ is not an independent function.
It is related to $x_c(\tau)$ because of the conservation of the total 
(dimensionless) planetesimal surface density
\begin{eqnarray}
M_1\equiv\int\limits_1^\infty xf(x)dx.
\end{eqnarray}
Taking this constraint into account one finds that
\begin{eqnarray}
\psi(\tau)=\left\{
\begin{array}{l}
x_c^{-2}(\tau),~~~~~~~~~~~~~\alpha<2,\\
x_c^{-\alpha}(\tau),~~~~~~~~~~~~~\alpha>2.
\end{array}
\right.
\label{eq:time_func}
\end{eqnarray}
In the case $\alpha>2$ mass spectrum behaves as $f(x,\tau)=x^{-\alpha}$
for $x\ll x_c(\tau)$; only the position of the high mass cutoff 
$x_c(\tau)$ shifts towards higher and higher masses with time.  

\begin{figure}
\plotone{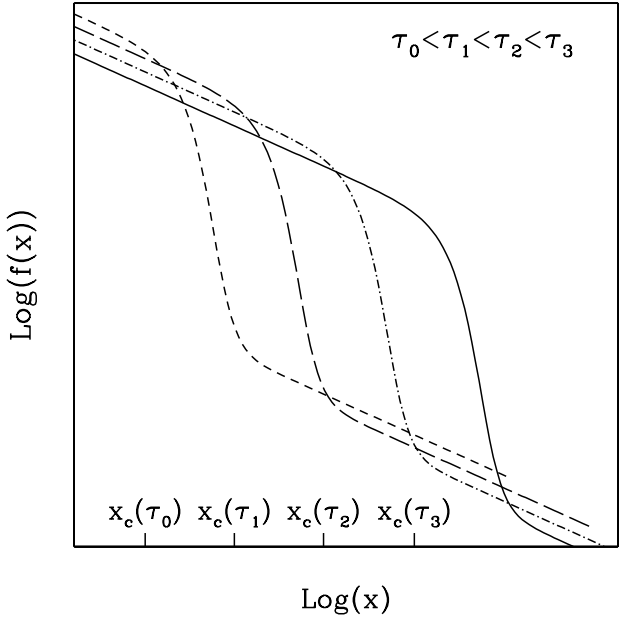}
\caption{Schematic representation of the shape 
of planetesimal mass spectrum 
and its evolution in time. Mass distribution
has a power law form exponentially cut off at $x_c(\tau)$ with a 
tail of runaway bodies for $x\gg x_c(\tau)$. 
\label{fig:mass}}
\end{figure}

We define $M_\nu(\tau)$ to be the $\nu$-th order moment of the mass distribution:
\begin{eqnarray}
M_\nu(\tau)\equiv\int\limits_1^\infty  (x^\star)^\nu f(x^\star,\tau)dx^\star.
\label{eq:M_def}
\end{eqnarray}
%Note that $M_1$ is a (dimensionless) total surface density of planetesimals
%in the disk, which is set constant in this study.
Integral in (\ref{eq:M_def}) is dominated by the upper end
of the power-law part of the mass 
spectrum (i.e. by $x^\star\sim x_c$) if $\nu>\alpha-1$;
in this case we can write [using (\ref{eq:s_sim})] 
\begin{eqnarray}
&& M_\nu(\tau)=\tilde M_\nu [x_c(\tau)]^{\nu+1}\psi(\tau)
\nonumber\\
&&
=\tilde M_\nu
\left\{
\begin{array}{l}
x_c^{\nu-1},~~~~~~\alpha<2,\\
x_c^{\nu+1-\alpha},~~~\alpha>2,
\end{array}
\right.,~~
\tilde M_\nu\equiv\int\limits_0^\infty  y^\nu \varphi(y)dy,
\label{eq:red_M_def}
\end{eqnarray}
where $\tilde M_\nu$ --- 
{\it reduced moment} of order $\nu$ --- is a time-independent constant
for a given mass distribution.

One of the interesting features exhibited by self-consistent coagulation
simulations is the development of the tail of high mass bodies beyond
the cutoff of the bulk distribution of planetesimals 
(Wetherill \& Stewart 1989; Inaba \etal 2001). This is interpreted as the
manifestation of the {\it runaway} phenomenon taking place in a coagulating 
system. To explore the possibility of the runaway scenario 
we add to the mass spectrum (\ref{eq:mass_spec}) a tail of high mass
planetesimals extending beyond the exponential cutoff $x_c$
(see Appendix \ref{app:numerical_procedure}). In doing this 
we always make sure that runaway tail contains a negligible part of 
the system's mass and does not affect its dynamical state (i.e. stirring
and friction caused by these high mass planetesimals are small) which 
is true if biggest bodies are not too massive (Rafikov 2003c). 
Under these assumptions the explicit functional form of the runaway 
tail is completely unimportant. Throughout this study, we use 
the term ``planetesimals'' for bodies belonging to the power-law part 
of the spectrum ($x\lesssim x_c$), and ``massive'' or ``runaway'' bodies
to denote the constituents of the tail ($x\gg x_c$).

We do not restrict the variety of possible functional dependences of 
$x_c$ on time $\tau$. It will turn out that all
our results can be expressed as some functions of 
$x_c$ which leads to the time-dependence of planetesimal 
velocities in a very general form. In 
some cases it will be important that 
$x_c(\tau)$ does not grow too fast, which, however, we believe
is a fairly weak constraint (see discussion in 
\S 4.1.2).

%Also, in some cases it will be useful to have a very sharp mass cutoff, then
%we assume 
%\begin{eqnarray}
%\varphi(y)= \left\{
%\begin{array}{l}
%y^{-\alpha},~~~~~~~~~~~~~~y\le 1,\\
%0,~~~~~~~~~~~~~~~~~~y> 1.
%\end{array}
%\right.
%\label{eq:spec_sharp}
%\end{eqnarray}
%Such a spectrum will be of use in exploring the limits of validity of the 
%assumptions made in this section.

%%%%%%%%%%%%%%%%%%%%%%%%%%%%%%%%%%%%%%%%%%%%%%%%%%%%%%%%%%%%%%%%%%%
%%%%%%%%%%%%%%%%%%%%%%%%%%%%%%%%%%%%%%%%%%%%%%%%%%%%%%%%%%%%%%%%%%%

\section{Velocity scaling laws.
\label{sect:scaling}}

To facilitate our treatment of planetesimal velocities we introduce a set of 
simplifications into our consideration:
\begin{itemize}  

\item We split mass spectrum into several regions, such that in each 
of them planetesimal interactions can be considered as dominated 
by a single dynamical regime (shear-dominated or dispersion-dominated).
Transitions between such regions are not considered but in principle 
might be treated by interpolation.

\item When studying the dispersion-dominated regime we neglect the difference 
between the vertical and horizontal velocity dispersions, and treat them 
by a single equation. We also set Coulomb logarithm equal to constant, because   
of its rather weak dependence on $s, s_z$ or $x$.

\end{itemize}
The validity and impact of these assumptions on the velocity spectrum 
are further checked using numerical techniques. Since we are mostly 
interested in qualitative behavior of solutions, 
we do not expect numerical constants to be correctly reproduced 
by our analysis; they can be trivially fixed using numerical calculations. 

Mass distributions are parametrized by the value
of power law exponent $\alpha$ [see (\ref{eq:mass_spec})]. 
%and the time dependence of the fiducial  
%planetesimal mass $x_c(\tau)$. 
We split our investigation according to the
value of $\alpha$ into $3$ cases: {\it shallow} mass spectrum,
$\alpha<2$, {\it intermediate} mass spectrum, $2<\alpha<3$,
and {\it steep} spectrum, $\alpha>3$ (this regime splits into two more 
important subcases, see \S 4.3). 
Note that for the sake of avoiding additional complications 
we do not consider borderline cases $\alpha=2, 3$; 
they can be easily studied in the framework of our approach 
if the need arises.

In our analytical work planetesimals are started 
with large enough $s$ and $s_z$
so that they interact with each other in  the 
dispersion-dominated regime. We then also assume that 
planetesimals with masses $x\lesssim x_c(\tau)$ 
(containing most of the mass) stay in the dispersion-dominated 
regime w.r.t. each other also at later time, i.e. 
\begin{eqnarray}
s(x,\tau),~s_z(x,\tau)\gg x^{1/3}~~~~\mbox{for}~~~~x\lesssim 
x_c(\tau).
\label{eq:condit}
\end{eqnarray}
This is a reasonable assumption for all mass spectra at 
the beginning of evolution, although for steep size 
distributions it may break down when maximum planetesimal  
mass becomes very large (see \S 4.3).

%%%%%%%%%%%%%%%%%%%%%%%%%%%%%%%%%%%%%%%%%%%%%%%%%%%%%%%%%%
%%%%%%%%%%%%%%%%%%%%%%%%%%%%%%%%%%%%%%%%%%%%%%%%%%%%%%%%%%

\subsection{Shallow mass spectrum.
\label{subsect:shallow}}

Planetesimal mass distributions shallower
than $m^{-2}$ result from coagulation 
of high-velocity planetesimals, when gravitational focusing
is unimportant and collision rate is determined by 
geometrical cross-section of colliding bodies 
(Wetherill \& Stewart 1993; Kenyon \& Luu 1998). It is worth 
remembering however that in highly dynamically excited disks
(1) energy dissipation in inelastic collisions must be important, 
see \S \ref{sect:dissipative}, and (2) planetesimal 
fragmentation cannot be ignored. Despite that, 
the case of shallow mass spectrum is interesting because 
it facilitates understanding of the disks with 
other planetesimal size distributions.

%%%%%%%%%%%%%%%%%%%%%%%%%%%%%%%%%%%%%%%%%%%%%%%%%%%%%%%%%%%%%%%%%%%

\subsubsection{Velocities of planetesimals.
\label{subsubsect:powlaw1}}

We start by considering planetesimals, $x\lesssim x_c(\tau)$ --- 
part of the size distribution containing most of the mass. 
Regarding them as 
interacting in the dispersion-dominated regime [i.e. the condition
(\ref{eq:condit}) is fulfilled] we may write
 using (\ref{eq:homog_heating_integr}) \& (\ref{eq:d-d}) that
\begin{eqnarray}
\frac{\partial s^2}{\partial \tau}=
\int\limits_1^{\infty}dx^\star f(x^\star)\frac{x^\star(x+x^\star)}
{s^2+s^{\star 2}}\nonumber\\
\times\left[\frac{x^\star}{x+x^\star} A_1
-2\frac{s^2}{s^2+ 
s^{\star 2}}A_2\right]
\label{eq:dd_eq}
\end{eqnarray}
for $x\ll x_c(\tau)$.
Since we use a single equation to describe the 
evolution of both $s$ and $s_z$, the values of constants $A_{1,2}$
are not well defined but this is not important for deriving general
properties of the velocity spectrum. 

\begin{figure}
\plotone{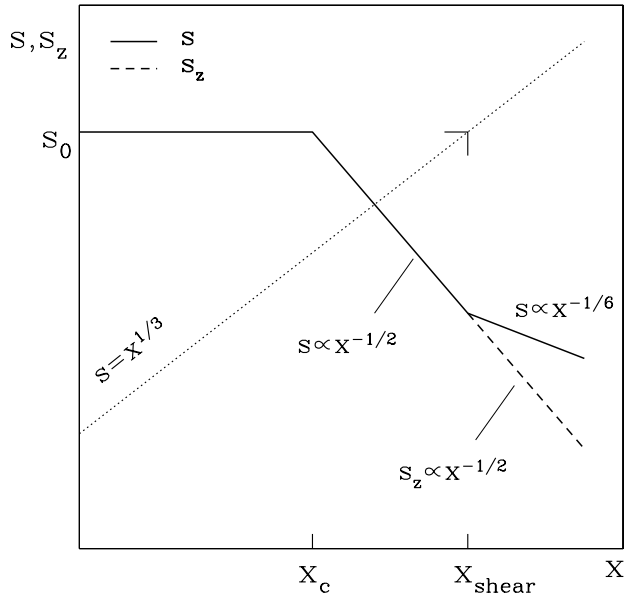}
\caption{
Sketch of the behavior of the scaled 
planetesimal eccentricity and inclination
$s=\sigma_e a/r_H$ and $s_z=\sigma_i a/r_H$ as
functions of their dimensionless mass $x=m/m_0$, predicted 
by asymptotic analysis of \S 4.1. 
Solid line displays $s(x)$ while dashed
line is for $s_z(x)$ (for $x<x_{shear}$ their behaviors are the same).
Dotted line delineates shear- and dispersion-dominated regimes
of planetesimal interaction with bodies of comparable mass. 
The rest of notation and theoretical predictions for the temporal 
behavior of $s(x)$ and $s_z(x)$ can be found in the text.
\label{fig:shallow}}
\end{figure}

We can represent (\ref{eq:dd_eq}) asymptotically in the following form:
\begin{eqnarray}
&& \frac{\partial s^2}{\partial \tau}\approx A_1 
\int\limits_1^{\sim x}dx^\star \frac{x^{\star 2} f(x^\star)}
{s^2+s^{\star 2}}-2A_2 s^2 x
\int\limits_1^{\sim x}dx^\star \frac{x^\star f(x^\star)}
{(s^2+s^{\star 2})^2}\nonumber\\&& +
\int\limits_{\sim x}^{\infty}dx^\star \frac{x^{\star 2} f(x^\star)}
{s^2+s^{\star 2}}
\left[A_1
-2\frac{s^2}{s^2+ 
s^{\star 2}}A_2\right].
\label{eq:dd_eq1}
\end{eqnarray}
The first two terms on the r.h.s. represent correspondingly the 
gravitational stirring and friction by bodies smaller than $x$.
The last term is responsible for the combined effect of stirring and 
friction produced by bodies bigger than $x$. From this equation it is 
easy to see that $s(\tau)$ independent of $x$ is a legitimate solution 
of (\ref{eq:dd_eq1}) in the case $\alpha<2$. 
Indeed, assuming $s$ to be independent of $x$
we find that all integrals over $x^\star$ in (\ref{eq:dd_eq1}) are 
dominated by their upper limits when mass spectrum is shallow. Using 
(\ref{eq:s_sim})-(\ref{eq:red_M_def}) we can estimate that
\begin{eqnarray}
&& \frac{\partial s^2}{\partial \tau}\approx A_1 
\frac{x_c}{2s^2}\left(\frac{x}{x_c}\right)^{3-\alpha}
-2 A_2 \frac{x_c}{4s^2}\left(\frac{x}{x_c}\right)^{3-\alpha}
\nonumber\\&& +(A_1-A_2)\frac{M_2(\tau)}{2s^2}.
\label{eq:dd_eq2}
\end{eqnarray}
Note that in deriving this expression we have extended
the integration range of the last term in
(\ref{eq:dd_eq1}) from $1$ to infinity (not from $\sim x$).
This is legitimate because this integral is dominated by 
the contribution coming from $\sim x_c\gg x$
and, thus, adding interval from $1$ to $\sim x$ to integration range 
would introduce only a subdominant contribution to the final answer. 
Clearly, two first terms in (\ref{eq:dd_eq2}) 
are negligible
%\footnote{In fact, it is 
%easy to see that 
%the first term is {\it always} smaller compared to third 
%if $\alpha<3$.} 
compared to the third one for $x\ll x_c$, and third term
is positive [see (\ref{eq:property})] and independent of 
$x$, meaning that $s(\tau)$ is also 
independent of mass, in accord with our assumption.
One can then easily find that\footnote{Throughout the 
paper we imply by $s_0$ the velocity dispersion of smallest 
planetesimals and $s_0$ is different for different mass spectra,
see (\ref{eq:vel_spec2}) and (\ref{eq:vel_spec4}) below.} 
[dropping constant 
factors $A_{1,2}$ but bearing in mind condition
(\ref{eq:property})]
\begin{eqnarray}
&& s(x,\tau)\approx s_0\equiv \left[\int\limits^\tau
M_2(\tau^\prime)d\tau^\prime \right]^{1/4}\nonumber\\
&& =
\left[\tilde M_2\int\limits^\tau
x_c(\tau^\prime)d\tau^\prime \right]^{1/4},~~
\alpha<2,~~x\lesssim x_c
\label{eq:vel_spec1}
\end{eqnarray}
[if $s(x,\tau)\gg s(x,0)$].
The second equality holds only for $\alpha<2$, according to 
(\ref{eq:time_func}) \& (\ref{eq:red_M_def}).  
Thus, in the case of mass spectrum  
shallower than $x^{-2}$ random velocities are
independent of planetesimal masses and uniformly grow with time. Both 
gravitational stirring and friction are dominated
by biggest planetesimals ($x\sim x_c$), with stirring being more 
important; friction by small bodies is completely negligible
and energy equipartition between planetesimals of 
different mass is not reached. For this reason, planetesimal velocity  
expressed in physical units using (\ref{eq:s_def}), 
(\ref{eq:tau}), \& (\ref{eq:vel_spec1}) 
is independent of the choice of $m_0$.
A schematic representation of  
velocity spectrum for the shallow mass distribution 
is displayed in Figure \ref{fig:shallow}.

%%%%%%%%%%%%%%%%%%%%%%%%%%%%%%%%%%%%%%%%%%%%%%%%%%%%%%%%%%%%%%%%%%%

\subsubsection{Velocities of runaway bodies.
\label{subsubsect:tail1}}

Now we turn our attention to the runaway bodies, 
$x\ga x_c$. By assumption, this tail 
contains so little mass, that it cannot affect velocities 
not only of planetesimals but also of runaway bodies
themselves. 
 
Because of our assumption of the dispersion-dominated 
interaction between 
all planetesimals with  $x\lesssim x_c$ it is natural to expect that
at least those runaway bodies which are not too heavy still experience 
dispersion-dominated scattering by planetesimals with masses 
$\lesssim x_c$. Assuming that the 
velocity dispersion of these runaway bodies is much smaller than 
that of the bulk of planetesimals --- natural consequence of 
the tendency to equipartition of random epicyclic energy 
between bodies of different mass ---
we can rewrite (\ref{eq:dd_eq}) as 
\begin{eqnarray}
\frac{\partial s^2}{\partial \tau}\approx A_1 
\int\limits_1^{\infty}dx^\star \frac{x^{\star 2} f(x^\star)}
{s^{\star 2}}-2A_2 s^2 x
\int\limits_1^{\infty}dx^\star \frac{x^\star f(x^\star)}
{s^{\star 4}}.
\label{eq:dd_eq_tail}
\end{eqnarray}
Although the integration range of all terms in the r.h.s. is 
extended to infinity, the exponential cutoff of $f(x)$ at
$\sim x_c$ effectively restricts the integration range to be from $1$
to $\sim x_c$. 

\begin{figure}
\epsscale{2.0}
\plottwo{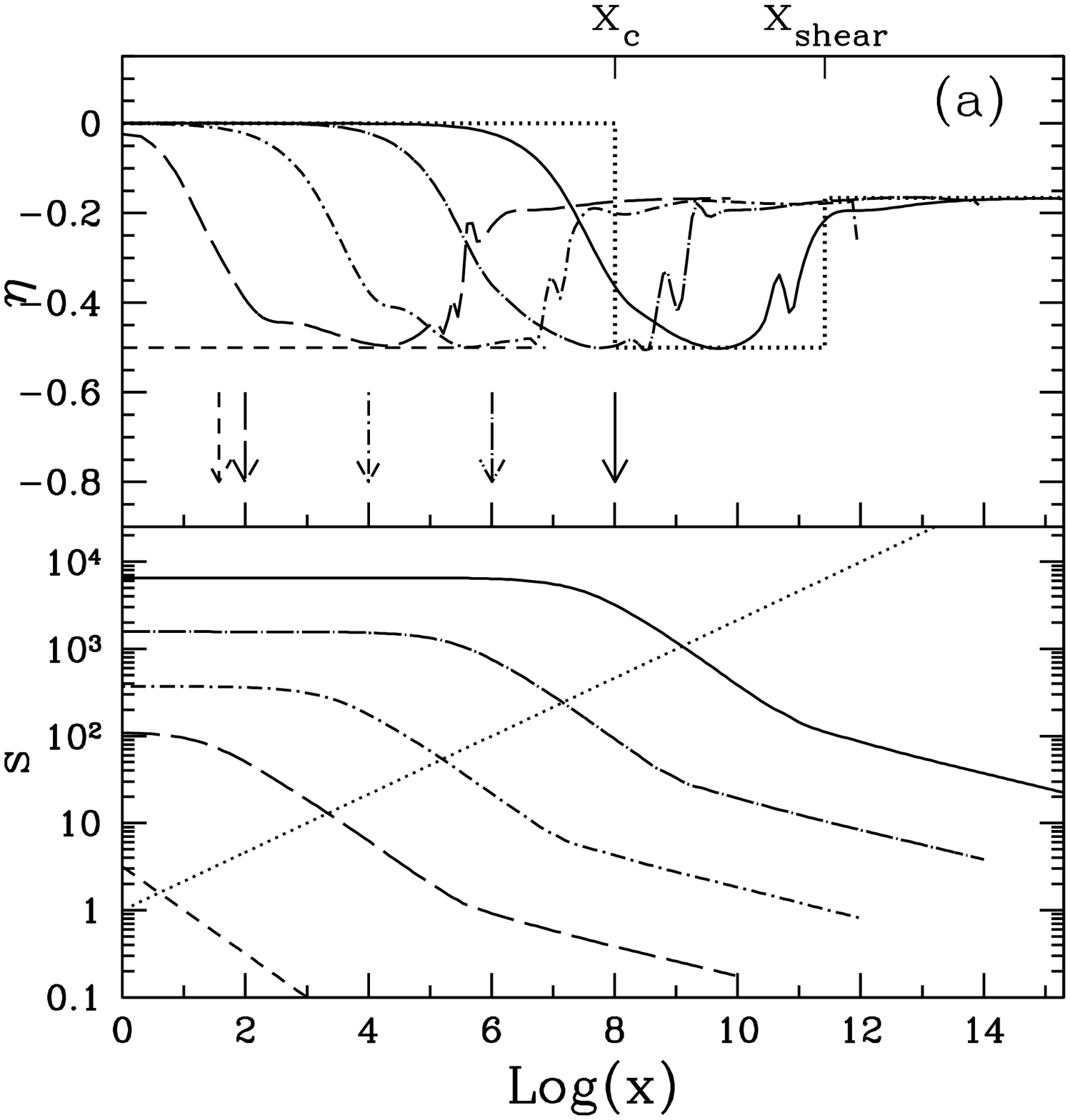}{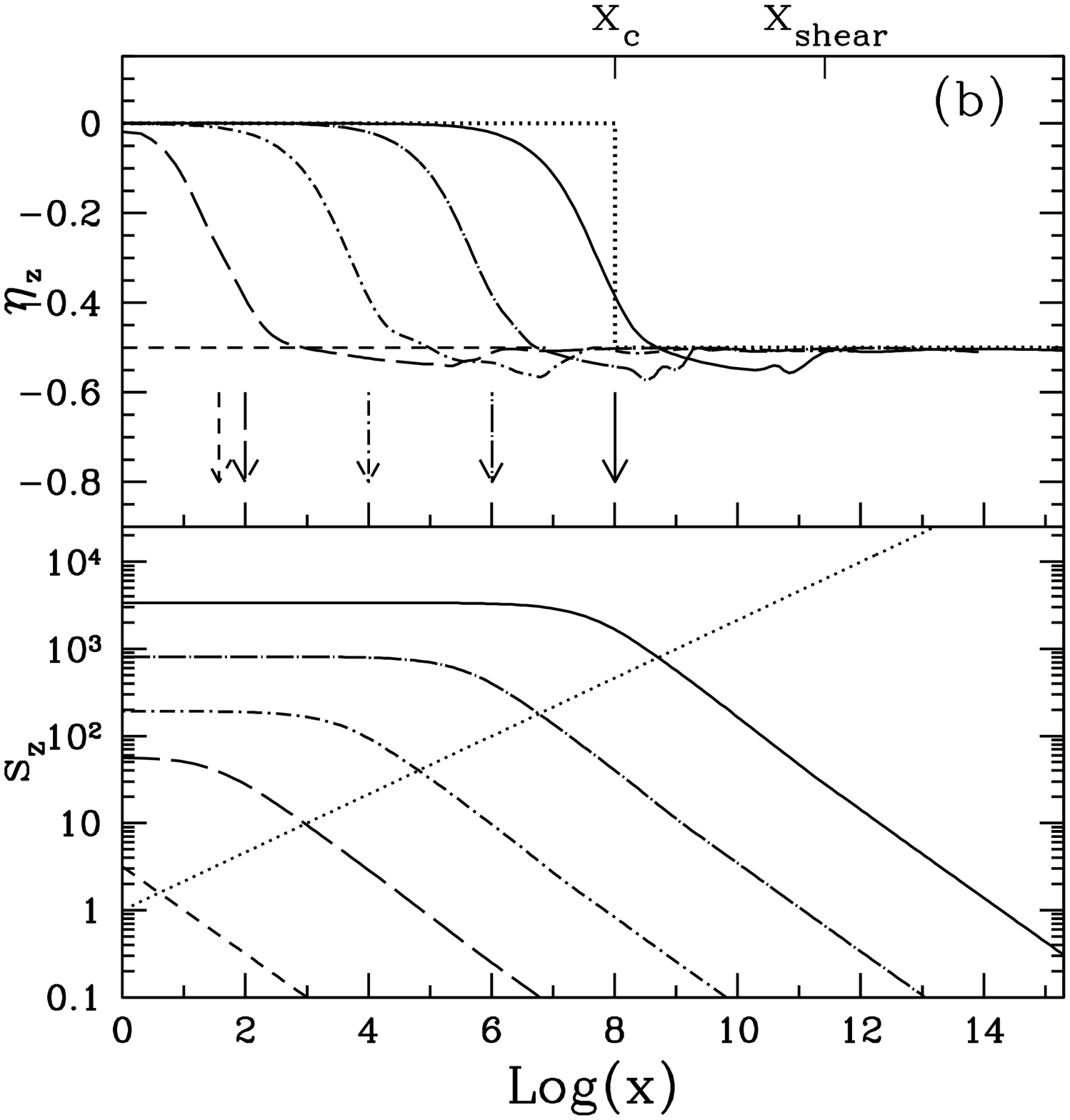}
\caption{
Numerical results in the case of $\alpha=1.5$ for (a) eccentricity 
dispersion (dimensionless) $s$ and (b) inclination dispersion $s_z$.
In the bottom panels the velocity spectrum is plotted as a function 
of dimensionless mass $x$, while in the top panels the power law
index $\eta=d\ln s/d\ln x$ of the spectrum is displayed. 
Different curves correspond to increasing values of
maximum planetesimal mass $x_c$ (and, consequently, time) 
indicated by arrows of corresponding line type on the upper panels. 
Dotted line in
the lower panels is $s=x^{1/3}$ (boundary between different velocity 
regimes). Dotted curve in the upper panels is the theoretical 
prediction for the run of the spectral index $\eta$ corresponding to 
the largest $x_c$ displayed (for which numerical result is always shown 
by thick solid line). These results are to be compared with 
theoretical predictions shown in Figure 
\ref{fig:shallow}.   
\label{fig:num_1.5}}
\end{figure}

One can easily see that $s(x,\tau)=const(x)$ cannot be a solution
of  (\ref{eq:dd_eq_tail}) because dynamical friction is too strong.
Indeed, if we were to assume the opposite, we would find that 
$s(x,\tau)$ is still given by (\ref{eq:vel_spec1}). However, 
direct substitution of (\ref{eq:vel_spec1}) into (\ref{eq:dd_eq_tail})
shows that the gravitational friction term exceeds other contributions
for $x\ga x_c$. This might tempt one to suggest that heating 
(first) term in r.h.s. (\ref{eq:dd_eq_tail}) should be neglected compared to
the friction (second) term. In this case 
one would find that $s(x,\tau)$ exponentially decays in time and very 
soon heating term becomes the dominant 
one, contrary to initial assumption. These results suggest 
 the only remaining possibility --- that heating of massive bodies 
by planetesimals almost exactly balances friction 
by planetesimals, and minute difference 
between them is accounted for by the time dependence of 
$s(x,\tau)$ embodied in the l.h.s. of (\ref{eq:dd_eq_tail}).
Given that this assumption is correct we immediately find that 
\begin{eqnarray}
&& s(x,\tau)\approx \nonumber\\
&& \frac{1}{\sqrt{x}}\left[
\int\limits_1^{\infty}dx^\star \frac{x^{\star 2} f(x^\star)}
{s^{\star 2}}\right]^{1/2}\left[
\int\limits_1^{\infty}dx^\star \frac{x^\star f(x^\star)}
{s^{\star 4}}\right]^{-1/2}.
\label{eq:vel_spec_tail}
\end{eqnarray}
It is easy to see that scaling $s\propto x^{-1/2}$
follows directly from assuming that (1) the interaction is
in the dispersion-dominated regime, and (2) heating of runaway 
bodies by planetesimals is balanced by friction also due to 
planetesimals. This specific result is completely independent of the 
mass or velocity spectra of small bodies. We will highlight this again 
when we obtain similar results for intermediate and steep mass spectra in 
\S 4.2.2 and \S C.2.
For a specific case $\alpha<2$ considered in this section 
$s^\star=s_0(\tau)$ and one finds that
\begin{eqnarray}
&& s(x,\tau)\approx  \frac{s_0(\tau)}{\sqrt{x}}\left[
\frac{M_2(\tau)}{M_1}\right]^{1/2}\nonumber\\
&& =
s_0\left(\frac{x_c}{x}\right)^{1/2}\left(
\frac{\tilde M_2}{M_1}\right)^{1/2},~~x_c\lesssim x
\lesssim x_{shear}
\label{eq:vel_spec_tail1}
\end{eqnarray}
[$x_{shear}$ is defined 
below in equation (\ref{eq:x_shear_def})].
At $x\sim x_c$ this formula is consistent with $s_0$ ---
extrapolation of small mass result (\ref{eq:vel_spec1}) 
up to $x\sim x_c$.

Time dependence enters (\ref{eq:vel_spec_tail1}) 
eventually through $x_c(\tau)$. Thus, it is the 
behavior of $x_c(\tau)$ that determines $\partial s^2/\partial\tau$
in (\ref{eq:dd_eq_tail}). The timescale on which $x_c$ varies 
is usually much longer than the dynamical relaxation time of the system, 
thus one would expect l.h.s. of (\ref{eq:dd_eq_tail}) to be 
small in accordance with the assumptions which led
to (\ref{eq:vel_spec_tail}). 
The requirement of the negligible effect of time derivatives
in situations like the one described here is the only
constraint which must be imposed on the possible behavior of $x_c(\tau)$.

%In principle, one can solve equation (\ref{eq:dd_eq_tail}) retaining all
%its terms:
%\begin{eqnarray}
%s(x,\tau)\approx\left\{C+\int\limits^\tau
%d\tau^\prime\frac{A_1 M_2(\tau^\prime)}{s_0^4(\tau^\prime)}
%\exp\left[2 A_1 M_1 x\int\limits^{\tau^\prime}
%\frac{d\tau^{\prime\prime}}{s_0^4(\tau^{\prime\prime})}\right]\right\}
%\exp\left[-2 A_1 M_1 x\int\limits^\tau
%\frac{d\tau^\prime}{s_0^4(\tau^\prime)}\right]
%\label{eq:est_eq}
%\end{eqnarray}
%Constant term inside the curly brackets is what one would obtain 
%keeping only the friction term in the r.h.s. of (\ref{eq:dd_eq_tail}). 
%One can easily see that, as we mentioned before,
%the contribution from this term quickly loses its importance. Dropping
%it completely and evaluating the remaining integral by parts one arrives at
%the equation (\ref{eq:vel_spec_tail1}) again (any additional 
%corrections to this result 
%have relative importance of order $x_c/x\ll 1$). Similar reasoning
%is in effect in all other cases where time derivatives in l.h.s. 
%are neglected (see e.g. \S \ref{subsubsect:powlaw2}).

As we consider even bigger $x$ we find that at some 
mass runaway bodies start interacting with bodies of 
similar mass in the shear-dominated regime; this happens when  
$x^{1/3}\sim s_0(x_c/x)^{1/2}$, or $x\sim x_c(s_0/x_c^{1/3})^{6/5}$. 
This, however does not affect the dynamics of these bodies,
because by our assumption runaway tail contains too little
mass to dynamically affect its own constituents. It is only 
after the runaway bodies start interacting in the shear-dominated 
regime  with {\it planetesimals} (i.e. bodies lighter than $x_c$),
that the velocity evolution of the tail gets affected.  
In our case this clearly happens when $x^{1/3}\sim s_0(\tau)$. Thus, 
runaway bodies with masses 
\begin{eqnarray}
x\gtrsim x_{shear}\equiv s_0^3
\label{eq:x_shear_def}
\end{eqnarray}
interact with all planetesimals in the 
shear-dominated regime. 

As mentioned in \S \ref{sect:velev}, 
in the shear-dominated case eccentricities and inclinations 
may no longer evolve along similar routes and we have to 
consider them separately.
Using (\ref{eq:homog_heating_integr}) \& (\ref{eq:shear}) we write 
\begin{eqnarray}
\frac{\partial s^2}{\partial \tau}=
\int\limits_1^{\infty}dx^\star f(x^\star)
x^\star(x+x^\star)^{1/3}\nonumber\\
\times\left[C_1\frac{x^\star}{x+x^\star}
-2C_2\frac{s^2}{(x+x^\star)^{2/3}}\right],\\
\frac{\partial s_z^2}{\partial \tau}=
\int\limits_1^{\infty}dx^\star f(x^\star)x^\star
\frac{s_z^2+s_z^{\star 2}}{(x+x^\star)^{1/3}}\nonumber\\
\times\left[D_1\frac{x^\star}{x+x^\star}
-2D_2\frac{s_z^2}{s_z^2+ 
s_z^{\star 2}}\right].
\label{eq:sd_eq}
\end{eqnarray}

%Assuming that $s,s_z\ll s^\star\approx 
%s_z^\star$ and $x\gg x^\star$ we 
%find that
%\begin{eqnarray}
%\frac{\partial s^2}{\partial \tau}=
%\frac{C_1}{x^{2/3}}
%\int\limits_1^{\infty}dx^\star x^{\star 2}f(x^\star)
%-
%2C_2\frac{s^2}{x^{1/3}}
%\int\limits_1^{\infty}dx^\star x^\star f(x^\star),
%\label{eq:sd_eq1}
%\\
%\frac{\partial s_z^2}{\partial \tau}=
%\frac{D_1}{x^{4/3}}
%\int\limits_1^{\infty}dx^\star x^{\star 2}
%s^{\star 2} f(x^\star)
%-
%2D_2\frac{s_z^2}{x^{1/3}}
%\int\limits_1^{\infty}dx^\star x^\star f(x^\star).
%\label{eq:sd_eqz1}
%\end{eqnarray}
%Neglecting l.h.s. of both equations (similar to the case
%$x_c\la x\la x_{shear}$) 
%we obtain the following result:

Assuming that $s,s_z\ll s^\star\approx 
s_z^\star$ (because of the gravitational friction) 
and $x\gg x^\star$ (both heating and cooling are dominated 
by planetesimals, not runaway bodies),
and neglecting l.h.s. of both equations (similar to the 
previously discussed situation for $x_c\la x\la x_{shear}$), 
we obtain the following result:
\begin{eqnarray}
&& s(x,\tau)\approx  x^{-1/6}\left[\frac{M_2(\tau)}
{M_1}\right]^{1/2},\nonumber\\
&& s_z(x,\tau)\approx \frac{1}{\sqrt{x}}
\left[\frac{1}{M_1}\int\limits_1^\infty
dx^\star x^{\star 2}
s^{\star 2} f(x^\star)\right]^{1/2}.
\label{eq:velz_spec_tail1.5}
\end{eqnarray}
Finally, setting $s^\star\approx s_0(\tau)$ as appropriate
for $\alpha<2$ we find that
\begin{eqnarray}
&& s(x,\tau)\approx 
x^{-1/6}\left(\frac{\tilde M_2}
{M_1}x_c\right)^{1/2},
\nonumber\\&&
s_z(x,\tau)\approx 
s_0\left(\frac{x_c}{x}\right)^{1/2}\left(
\frac{\tilde M_2}{M_1}\right)^{1/2},~~x\ga x_{shear}.
\label{eq:velz_spec_tail2}
\end{eqnarray}

\begin{figure}
\epsscale{1.27}
\plotone{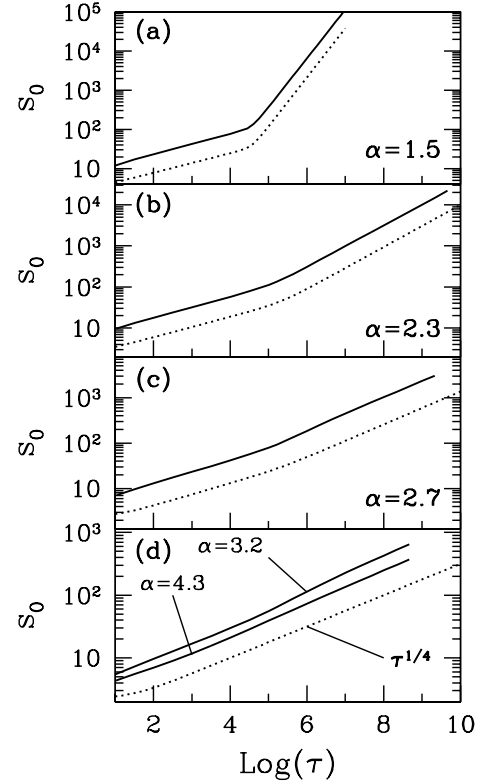}
\caption{
Comparison of numerical ({\it solid curves}) and analytical 
({\it dotted curves}) behaviors of $s_0$ as a function of 
dimensionless time $\tau$. We display cases: (a) $\alpha=1.5$, 
(b) $\alpha=2.3$, (c) $\alpha=2.7$, (d) $\alpha=3.2$ and $\alpha=4.3$. 
\label{fig:s_0}}
\end{figure}

These expressions demonstrate that transition to the shear-dominated 
regime is accompanied by a change of the power-law index of 
eccentricity scaling with mass, while the power-law slope of inclination
scaling stays the same. This happens 
because in the case of inclination evolution heating term is functionally 
similar to the friction term 
(exactly like in the dispersion-dominated case), which is not the case 
for eccentricity evolution, see (\ref{eq:shear}).
As a result, for $x\la x_{shear}$ one finds that $s\gg s_z$ --- planetesimal 
velocity ellipsoid flattens considerably in the vertical 
direction, which is very unlike the dispersion-dominated case 
(see Figure \ref{fig:shallow}).
This conclusion is very general and should be found whenever 
runaway bodies interact with {\it all} planetesimals in 
the shear-dominated regime.

%%%%%%%%%%%%%%%%%%%%%%%%%%%%%%%%%%%%%%%%%%%%%%%%%%%%%%%%%%%%%%%%%%%

\subsubsection{Comparison with numerical results.
\label{subsubsect:numer1}}

To check the prediction of our asymptotic analysis we have performed
numerical calculations of planetesimal velocity evolution
for a distribution of planetesimal masses which was artificially 
evolved in time according to prescriptions outlined in 
\S \ref{sect:mass_spec}. Details of our numerical procedure can be found in 
Appendix \ref{app:numerical_procedure}.

In Figure \ref{fig:num_1.5} we
show the evolution of planetesimal eccentricity and inclination 
dispersions for a shallow mass spectrum with $\alpha=1.5$. For each 
curve displaying $s$ or $s_z$ run with $x$ at a specific moment of time
we also computed power law spectral index $\eta\equiv d\ln s/d\ln x$.
We then compare power law indices $\eta$ and $\eta_z$ with analytical
predictions given by equations  (\ref{eq:vel_spec1}), 
(\ref{eq:vel_spec_tail1}), \& (\ref{eq:velz_spec_tail2}).
Dotted line in the upper panels of Figure \ref{fig:num_1.5} represents
analytical predictions for $\eta$ and $\eta_z$ at the moment when the 
snapshot of velocity spectrum displayed by the solid line
(corresponding to the highest $x_c$ shown) was taken. 
One can see that agreement between two approaches is quite good. 
Power law slopes of both eccentricity $\eta$ and inclination $\eta_z$ 
exhibit a well defined plateau at masses considerably smaller than 
$x_c$ which is predicted by (\ref{eq:vel_spec1}). Above upper mass 
cutoff $\eta$ and $\eta_z$ plunge towards
$-1/2$ which is in agreement with (\ref{eq:vel_spec_tail1}), although 
they do not follow this value very closely. The reason is most likely 
the lack of dynamical range --- transitions between different regimes 
typically occupy substantial intervals in mass. Finally, at 
$x$ above $x_{shear}$, where runaway bodies interact with all 
planetesimals in shear-dominated regime, $\eta$ goes up to $-1/6$
while $\eta_z$ stays at a value of $-1/2$, exactly like 
equation (\ref{eq:velz_spec_tail2}) predicts. Glitches in the 
curves of $\eta$ and $\eta_z$ in the vicinity of $x_{shear}$
are the artifacts of interpolation of 
scattering coefficients between shear- and 
dispersion-dominated regimes.

In Figure \ref{fig:s_0}a we display the time evolution of $s_0$ ---
eccentricity dispersion of smallest planetesimals versus the prediction
of our analysis given by equation (\ref{eq:vel_spec1}). One can see that
the agreement between two approaches is excellent --- analytical and 
numerical results  closely 
follow each other as time increases. Vertical offcet 
between them amounts to a constant
factor by which numerically determined $s_0(\tau)$ differs from analytical 
$s_0(\tau)$. This, of course, is expected because  
we completely ignored constant coefficients in our 
asymptotic analysis. Numerical
results shown in Figure \ref{fig:s_0}a allow us to fix constant
coefficient in (\ref{eq:vel_spec1}), and we do this when we calculate
$x_{shear}$ in Figure \ref{fig:num_1.5} using equation 
(\ref{eq:x_shear_def}). Note that the rapid growth of $s_0$ with
time should certainly increase 
velocities of planetesimals beyond their escape speeds 
(if it were not the case initially). Dynamical evolution in this case
is discussed in more details in \S \ref{sect:dissipative}.

An apparent break in the behavior of $s_0(\tau)$ at $\tau\sim 10^5$
can be seen in Figure \ref{fig:s_0}a. It occurs because this is the 
time at which mass scale $x_c$ starts to grow. The prescription
for $x_c(\tau)$ which we use (described in Appendix 
\ref{app:numerical_procedure}) is such that $x_c$ is almost constant
for $\tau\lesssim 10^5$; as a result, $s_0\propto \tau^{1/4}$ 
according to equation (\ref{eq:vel_spec1}). At $\tau\gtrsim 10^5$ 
scale factor starts to increase rapidly and this causes $s_0$ to
grow faster. To conclude, the results of this section show 
that the overall agreement between numerical calculations and 
analytical theory of planetesimal velocity evolution is rather good.

%%%%%%%%%%%%%%%%%%%%%%%%%%%%%%%%%%%%%%%%%%%%%%%%%%%%%%%%%%%%%%%%%%%
%%%%%%%%%%%%%%%%%%%%%%%%%%%%%%%%%%%%%%%%%%%%%%%%%%%%%%%%%%%%%%%%%%%

\subsection{Intermediate mass spectrum.
\label{subsect:interm}}

Self-consistent coagulation calculations (Kenyon \& Luu 1998) 
and N-body simulations of planet formation (Kokubo \& Ida 1996, 2000)
often yield intermediate mass distributions of planetesimals 
($2<\alpha<3$). Kokubo \& Ida (1996) have found, for example 
$\alpha\approx 2.5\pm 0.4$ in their N-body calculation of
collisional evolution of several thousand massive ($m=10^{23}$ g)
planetesimals. Such outcome seems to be ubiquitous for planetesimals
which agglomerate under the action of strong gravitational focusing 
(planetesimal velocities are well below the escape velocities from their 
surfaces) in the dispersion-dominated regime, which makes the
case of intermediate mass spectra very important.

\begin{figure}
\epsscale{1.0}
\plotone{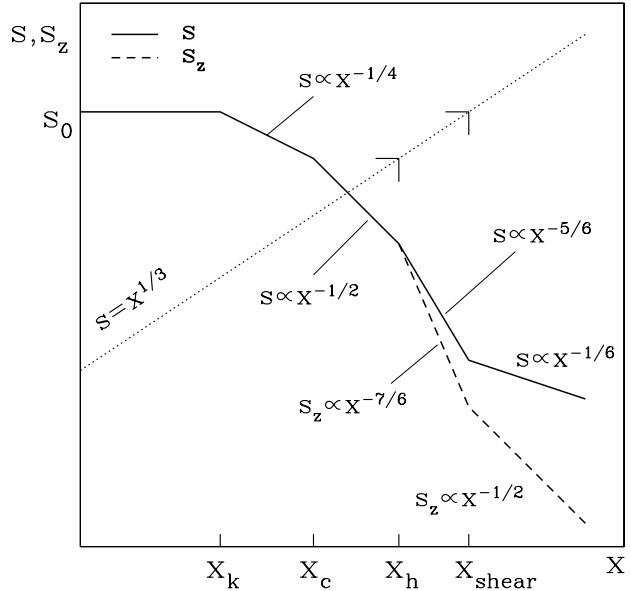}
\caption{
Same as Figure \ref{fig:shallow} but for the case of intermediate 
mass spectrum. Specific case of $2<\alpha<5/2$ is displayed.
Notation is explained in the text. Analogous plot for $5/2\le \alpha<3$
would exhibit different behavior of $s_z$ in the high-mass tail and
can be easily constructed using the results of \S 4.2.2.
\label{fig:interm}}
\end{figure}

%%%%%%%%%%%%%%%%%%%%%%%%%%%%%%%%%%%%%%%%%%%%%%%%%%%%%%%%%%%%%%%%%%%

\subsubsection{Velocities of planetesimals.
\label{subsubsect:powlaw2}}

As in \S 4.1 we start our consideration 
of mass spectra with $2<\alpha<3$ by concentrating on
planetesimals with masses $\lesssim x_c(\tau)$, assuming that
all these bodies interact in the dispersion-dominated regime 
(see \S \ref{sect:scaling}). Then equations 
(\ref{eq:dd_eq}) and  (\ref{eq:dd_eq1}) continue to hold.

It is natural to start by assuming again that $s$ is independent of $x$.
Performing analysis similar to that of 
\S 4.1.2 we find that 
stirring is still dominated by large planetesimals ($\sim x_c$), but
the biggest contribution to the gravitational friction is produced by 
smallest planetesimals (with $m\sim m_0$, or $x\sim 1$ in our 
dimensionless notation): 
\begin{eqnarray}
\frac{\partial s^2}{\partial \tau}\approx A_1 
\frac{x^{3-\alpha}}{2s^2}
-2 A_2 \frac{x}{4s^2}M_1+
(A_1-A_2)\frac{M_2(\tau)}{2s^2}.
\label{eq:dd_eq3}
\end{eqnarray}
Clearly, the first term on the r.h.s. which is responsible for the stirring
by small planetesimals is negligible compared to the last one, representing 
stirring by planetesimals bigger than $x$ [see (\ref{eq:red_M_def})]. 
Also, by definition, 
$M_1$ is just a dimensionless total surface density
of planetesimal disk, and thus has to be constant in time. 
Then our assumption of $s$ independent of $x$ is 
self-consistent and $s$ is given by the first equality 
in (\ref{eq:vel_spec1}) only if friction by small planetesimals 
[ second term in (\ref{eq:dd_eq3})] 
is small compared to stirring by big ones (third term), 
which is true provided that
\begin{eqnarray}
x\lesssim x_k(\tau) \equiv\frac{M_2(\tau)}{M_1}=
x_c\frac{\tilde M_2}{M_1}x_c^{2-\alpha}\ll x_c.
\label{eq:small_cond}
\end{eqnarray}
The last equality follows from the fact that 
$M_2=\tilde M_2 x_c^{3-\alpha}$ when $2<\alpha<3$. Thus, for small 
planetesimals ($x\lesssim x_k$) friction by even smaller 
ones is again unimportant;
their velocities are excited by big planetesimals ($x\sim x_c$), 
and as a result [cf. (\ref{eq:vel_spec1})]
\begin{eqnarray}
&& s(x,\tau)\approx s_0(\tau)= 
\left[\tilde M_2\int\limits^\tau
[x_c(\tau^\prime)]^{3-\alpha}
d\tau^\prime \right]^{1/4}\nonumber\\
&& =const(x),~~2<\alpha<3,~~~
x\lesssim x_k.
\label{eq:vel_spec2}
\end{eqnarray}

However, for $x\gtrsim x_k$, gravitational friction by small planetesimals
is no longer negligible compared to the velocity excitation by big ones. 
%In this case we make another assumption ---
%that the dynamical friction by small bodies balances excitation 
%by big ones. 
One would then expect friction to lower the 
velocities of big
bodies below those of small planetesimals (because of the tendency 
to energy equipartition associated with friction term). 
Starting with (\ref{eq:dd_eq1}) and using
(\ref{eq:vel_spec2}) we obtain the 
following equation for $x\gtrsim x_k$:
\begin{eqnarray}
&& \frac{\partial s^2}{\partial \tau}=
\frac{A_1}{s_0^2}
\int\limits_1^{\sim x_k}dx^\star (x^\star)^2 f(x^\star)
+
A_1\int\limits_{\sim x_k}^{\sim x}dx^\star \frac{(x^\star)^2 f(x^\star)}{s^{\star 2}}
\nonumber\\&& -
2A_2\frac{s^2x}{s_0^4}\int\limits_1^{\sim x_k}dx^\star x^\star f(x^\star)
-
2A_2 s^2x\int\limits_{\sim x_k}^{\sim x}dx^\star \frac{x^\star f(x^\star)}{s^{\star 4}}
\nonumber\\&& +
\frac{A_1-2A_2}{s^2}\int\limits_{\sim x}^{\infty}dx^\star (x^{\star})^2f(x^\star).
\label{eq:dd_eq4}
\end{eqnarray}
In this equation different terms represent stirring or friction by different 
parts of planetesimal mass spectrum, below and above $x_k$.
We now take an educated guess that the solution for $s$ can be 
obtained by equating third and fifth terms in the r.h.s. of (\ref{eq:dd_eq4}).
Physically this means that velocity spectrum results from the 
balance of velocity stirring by biggest planetesimals (masses $\sim x_c\gg x$,
similar to the case of shallow mass spectrum)
and friction which is mainly contributed by 
smallest planetesimals. One then finds that [see (\ref{eq:small_cond})]
\begin{eqnarray}
&& s(x,\tau)\approx
s_0(\tau)\left[
\frac{M_2(\tau)}{M_1}\right]^{1/4} x^{-1/4}\nonumber\\
&& \approx s_0(\tau)
\left[\frac{x_k(\tau)}{x}\right]^{1/4},~~x_k\lesssim x\lesssim x_c.
\label{eq:vel_spec3}
\end{eqnarray}
Substituting this solution into (\ref{eq:dd_eq4}) 
we easily verify that our neglect of all other terms in 
the r.h.s. of (\ref{eq:dd_eq4}) is indeed justifiable. 
The l.h.s. of (\ref{eq:dd_eq4}) can be neglected 
for $x_k\lesssim x\lesssim x_c$ on the
basis of arguments identical to 
those advanced in \S 4.1.2.
Thus, $s$ given by (\ref{eq:vel_spec3}) is indeed a legitimate 
solution in this mass range.

\begin{figure}
\epsscale{2.0}
\plottwo{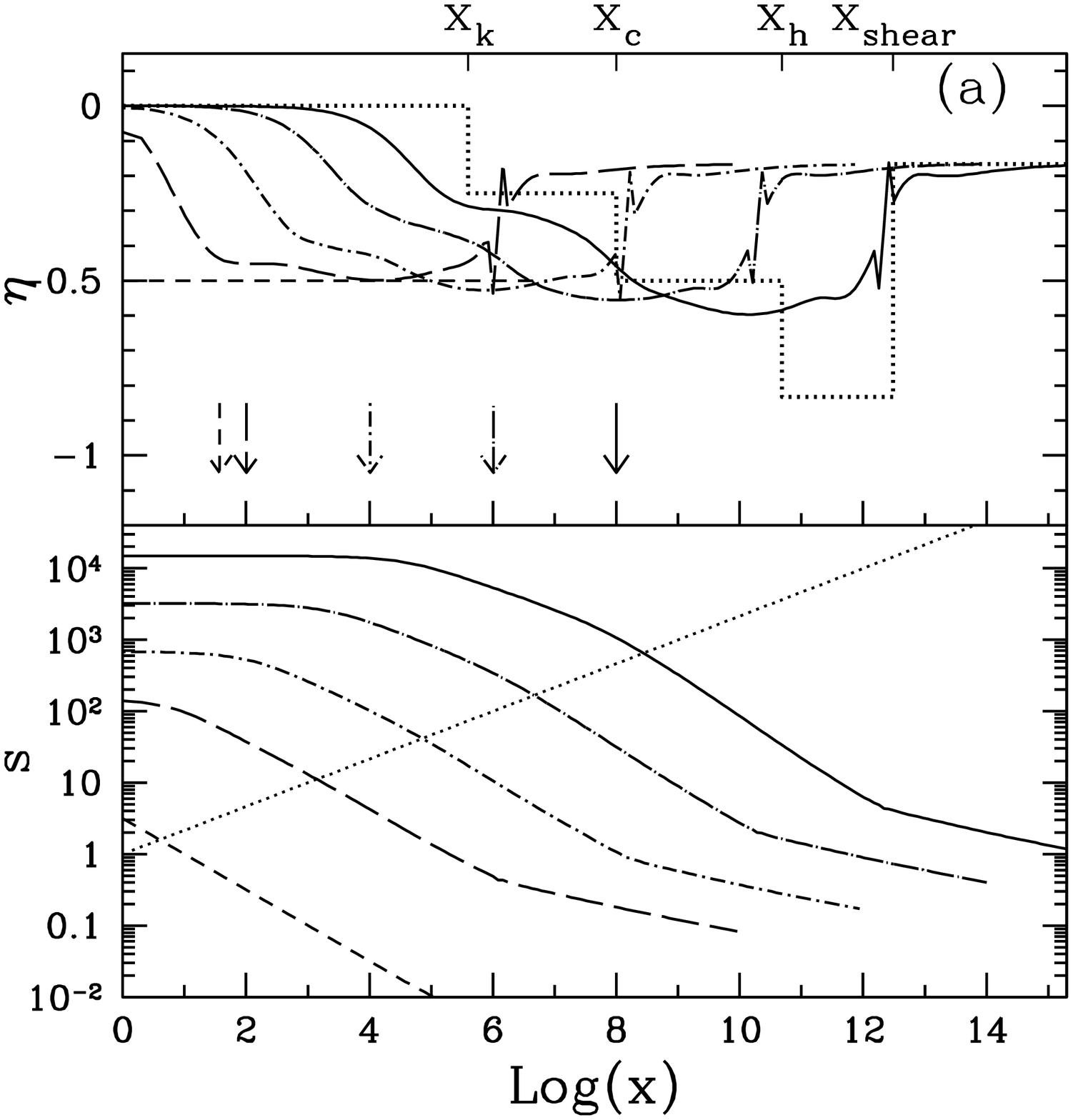}{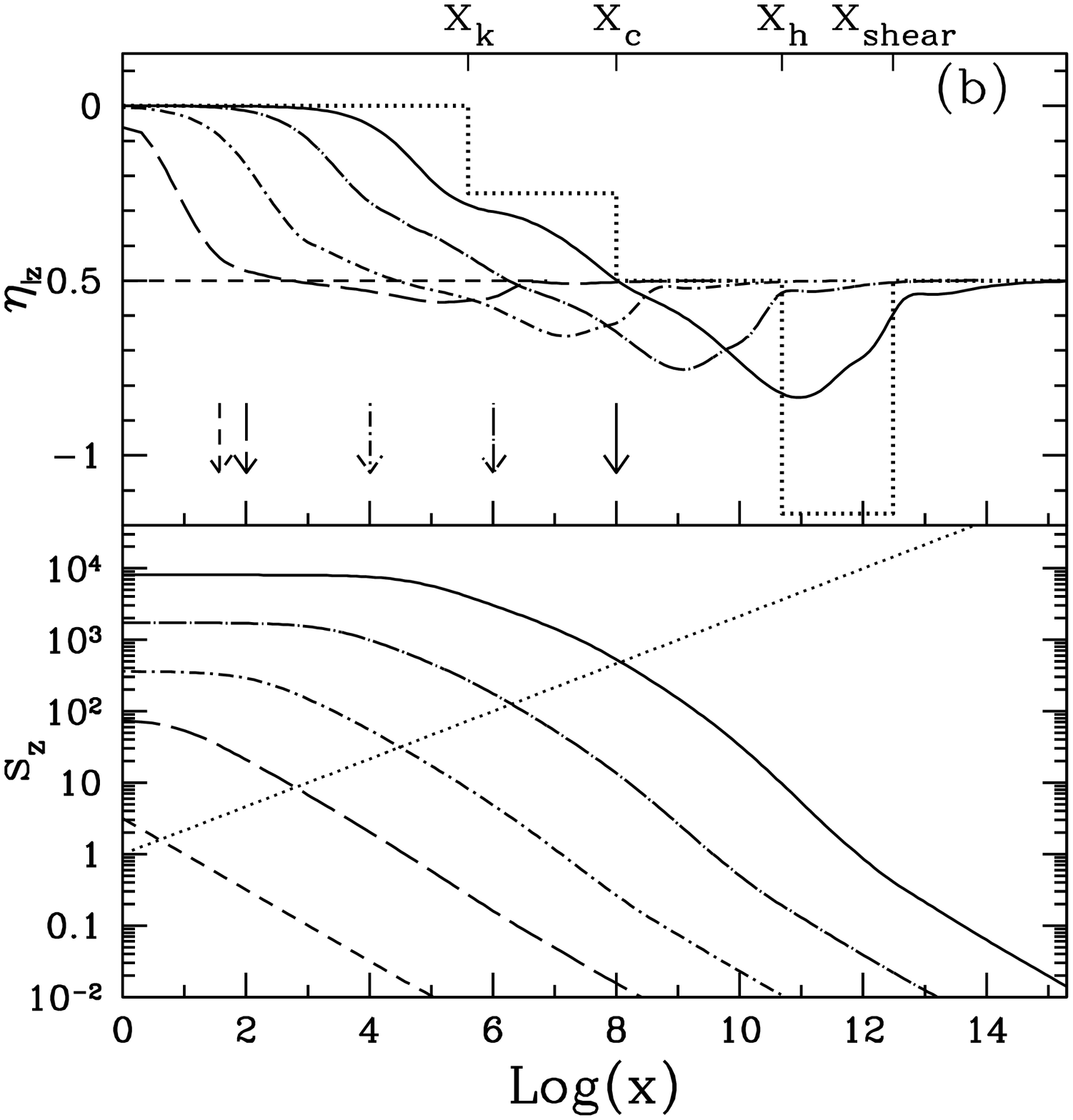}
\caption{
The same as Figure \ref{fig:num_1.5} but for the case of intermediate 
mass spectrum with  $\alpha=2.3$ (cf. Fig. \ref{fig:interm}).
\label{fig:num_2.3}}
\end{figure}

Equations (\ref{eq:vel_spec2}) \& (\ref{eq:vel_spec3}) highlight an interesting
feature of the case $2<\alpha<3$: although the mass spectrum
has a power law form with a continuous slope all the way up to $x_c$, velocity 
spectrum exhibits a knee at some intermediate mass $x_k$ (see Figure \ref{fig:interm}).
The resemblance of the velocity spectrum for $x\lesssim x_k$ to the 
spectrum for $\alpha<2$ is due to the fact that in both cases stirring is done 
by biggest planetesimals and friction has a small effect. However, for 
$x\gtrsim x_k$ effect of friction becomes important, and now it is dominated by
the part of mass spectrum (smallest planetesimals) different from that 
responsible for stirring (biggest planetesimals). This causes velocity spectrum 
to change its slope from $0$ to $-1/4$ in this mass range
(see Figure \ref{fig:interm}).

%%%%%%%%%%%%%%%%%%%%%%%%%%%%%%%%%%%%%%%%%%%%%%%%%%%%%%%%%%%%%%%%%%%

\subsubsection{Velocities of runaway bodies.
\label{subsubsect:tail2}}

Similar to the case of shallow mass spectrum, some (lightest)
runaway bodies interact with the bulk of
planetesimals in the dispersion-dominated regime. In this case
all the considerations of \S 4.1.2 hold, i.e. 
evolution equation (\ref{eq:dd_eq_tail}) stays unchanged and after 
the same line of arguments we arrive at the expression 
(\ref{eq:vel_spec_tail}) for the planetesimal velocity dispersion.
Integrals entering (\ref{eq:vel_spec_tail}) must be reevaluated 
anew using equations (\ref{eq:vel_spec2}) \& (\ref{eq:vel_spec3})
appropriate for $2<\alpha<3$. 
Performing this procedure and substituting resulting expressions 
into (\ref{eq:vel_spec_tail}) we find that
\begin{eqnarray}
&& s(x,\tau)\approx\frac{s_0}{\sqrt{x}}\left[
\frac{M_{5/2}(\tau)}{M_1 x_k^{1/2}(\tau)}\right]^{1/2}
\nonumber\\ && =
s_0\left(\frac{x_c}{x}\right)^{1/2}
\left(\frac{\tilde M_{5/2}}{\tilde M_2}
\sqrt{\frac{x_k}{x_c}}\right)^{1/2},~~
x_c\lesssim x\lesssim x_h.
\label{eq:vel_spec_tail3}
\end{eqnarray}
This velocity distribution is valid only for runaway
bodies lighter than
\begin{eqnarray}
x_h\equiv \left[s(x_c)\right]^3\approx 
s_0^3\left(\frac{x_k}{x_c}\right)^{3/4},
\label{eq:x1def}
\end{eqnarray}
because only such bodies interact with {\it all} small mass 
planetesimals ($x\lesssim x_c$) in the dispersion-dominated regime.
In agreement with what we found
earlier in \S 4.1.2, velocity dispersion decreases with 
mass as $x^{-1/2}$. At $x\sim x_c$ this expression matches the 
low mass result (\ref{eq:vel_spec3}). 

Runaway bodies heavier than $x_h$ but still lighter than 
$x_{shear}\equiv s_0^3$ are in a mixed state: 
they interact with the most massive  
planetesimals in the 
shear-dominated regime, and with lighter ones in the 
dispersion-dominated regime. This, of course, is a direct consequence 
of the fact that the velocity spectrum of planetesimals has a knee 
at $x_k$, see \S 4.2.1. 
Introducing 
\begin{eqnarray}
x_s(x)\equiv x_k\left(\frac{s_0}{x^{1/3}}\right)^4.
\label{eq:xsx_def}
\end{eqnarray}
we find that runaway bodies with masses $x$ between 
$x_h$ and $x_{shear}$
interact with planetesimals less massive than 
$x_s(x)$ in the dispersion-dominated regime, and with 
planetesimals heavier than $x_s(x)$ in the shear-dominated regime. 
Self-consistent analysis of planetesimal velocity spectrum for 
$x_h\lesssim x\lesssim x_{shear}$ is described in detail in 
Appendix \ref{app:subsubsect:tail2}, and only final results are shown here.
It turns out that eccentricity scales as 
\begin{eqnarray}
&& s(x,\tau)\approx \frac{s_0}{x^{5/6}}\left[
s_0^2\frac{M_{2}(\tau)}{M_1}\right]^{1/2}=
s_0\frac{s_0 x_k^{1/2}}{x^{5/6}},\nonumber\\
&& ~~~~~~~~~~~~~~~
~~~~~~~~~~~~~~~~~~~x_h\lesssim x\lesssim x_{shear},
\label{eq:vel_spec_tail4}
\end{eqnarray}
while inclination behaves in a different manner:
\begin{eqnarray}
&& s_{z}(x,\tau)\approx
\frac{s_0}{x^{7/6}}\left[
s_0^4 x_k^{1/2}\frac{M_{3/2}(\tau)}{M_1}\right]^{1/2}\nonumber\\
&& =\frac{s_0}{x^{7/6}}\left[s_0^4 x_k\left(
\frac{x_k}{x_c}\right)^{1/2}\frac{\tilde M_{3/2}}{\tilde M_2}
\right]^{1/2},~~~\alpha<5/2,
\label{eq:vel_spec_tailz4.5}\\
&& s_{z}(x,\tau)\approx s_0\left[
\frac{(x_s(x))^{7/2-\alpha}}{x_k^{1/2} x M_1}\right]^{1/2}
\nonumber\\
&& \propto x^{(4\alpha-17)/6},
~~~~~~~~~~~~~~~~~~~~~~~~~~~~~~\alpha>5/2,
\label{eq:vel_spec_tailz5}
\end{eqnarray}
In the case of $\alpha=5/2$ a more general expression for 
$s_z$ following directly from (\ref{eq:hybridz1}) 
should be used.

\begin{figure}
\epsscale{2.0}
\plottwo{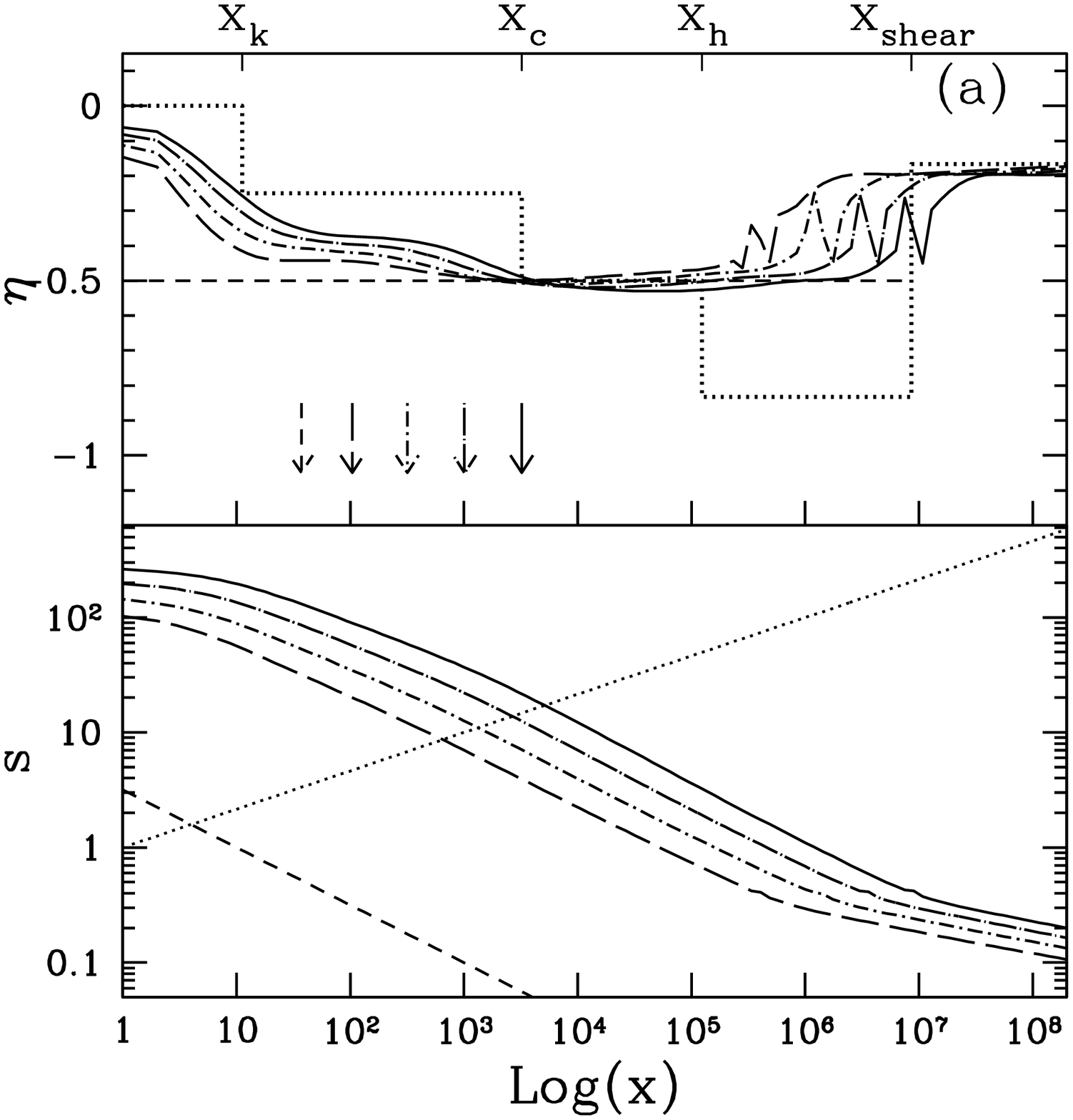}{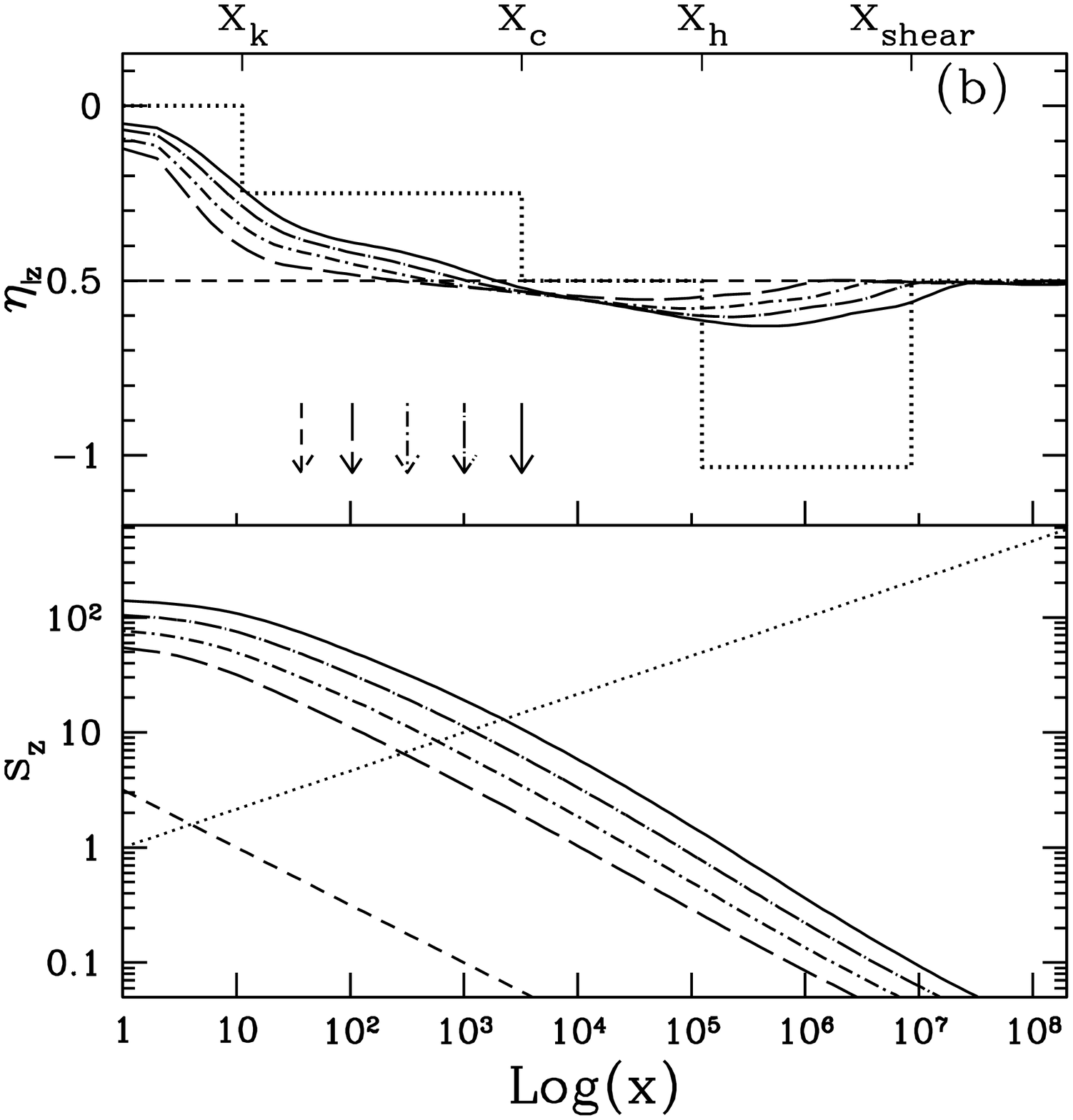}
\caption{
The same as Figure \ref{fig:num_1.5} but for the case of intermediate 
mass spectrum with  $\alpha=2.7$ (cf. Fig. \ref{fig:interm}).
\label{fig:num_2.7}}
\end{figure}

Finally, bodies heavier than $x_{shear}$ interact with
{\it all} planetesimals in the shear-dominated 
regime. Expressions (\ref{eq:velz_spec_tail1.5}) 
adequately describe velocity behavior in this case. 
Manipulating them with the aid of equations (\ref{eq:vel_spec2}) 
and (\ref{eq:vel_spec3}) we find that
\begin{eqnarray}
s(x,\tau)\approx\frac{x_k^{1/2}}{x^{1/6}},~~~~~~~~~~~~
x\ga x_{shear}.
\label{eq:vel_spec_tail6}
\end{eqnarray}
while 
\begin{eqnarray}
&& s_z(x,\tau)\approx s_0\left(\frac{x_k}{x}\right)^{1/2}
\left[\left(\frac{x_k}{x_c}\right)^{1/2}
\frac{\tilde M_{3/2}}{\tilde M_2}\right]^{1/2},\nonumber\\
&& ~~~~~~~~~~~~~~~~~~~~~~~~~~
~~~~~~~~~~~~~~~~~~~~~~\alpha<5/2,
\label{eq:vel_spec_tailz6}
\\
&& s_z(x,\tau)\approx s_0\left(\frac{x_k}{x}\right)^{1/2}
\left(\frac{x_k}{x_c}\right)^{(3-\alpha)/2},
~\alpha>5/2.
\label{eq:vel_spec_tailz7}
\end{eqnarray}
Scalings of velocity dispersions with mass $x$ 
in (\ref{eq:vel_spec_tail6})-(\ref{eq:vel_spec_tailz7})
are the same as in the case of shallow mass spectrum, but the time 
dependences are different. Velocity behavior in the case of  
intermediate mass spectrum is displayed in Figure \ref{fig:interm}.

%%%%%%%%%%%%%%%%%%%%%%%%%%%%%%%%%%%%%%%%%%%%%%%%%%%%%%%%%%%%%%%%%%%

\subsubsection{Comparison with numerical results.
\label{subsubsect:numer2}}

We calculated numerically the evolution of $s$ and $s_z$ for two
intermediate mass spectra: $\alpha=2.3$ and $\alpha=2.7$.  
In Figure \ref{fig:s_0}b,c we compare numerically
calculated $s_0(\tau)$ --- eccentricity dispersion of smallest 
planetesimals --- with predictions of our asymptotic theory. As in
the case of shallow mass spectrum discussed in 
\S 4.1.3, agreement between the two approaches is 
very nice. This allows us to fix coefficient in
(\ref{eq:vel_spec2}), absence of which in our asymptotic analysis 
causes a constant offset between numerical and analytical curves in  
Figure \ref{fig:s_0}b,c. Results
for the velocity dependencies on mass at different moments of time 
and their comparison with analytical theory of intermediate mass 
spectra outlined above are shown in Figures \ref{fig:num_2.3} and 
\ref{fig:num_2.7}.

In the case $\alpha=2.3$ one can clearly see the appearance of all
features we described in \S 4.2.1 and 
\S 4.2.2. Indeed, in Figure \ref{fig:num_2.3}
one can observe both the zero-slope part of the spectrum at 
$x\lesssim x_k$ and the trend of reaching the slope $-1/4$ for
$x_k\lesssim x \lesssim x_c$; the last feature is not fully
developed when we stop our calculation but it is almost certain 
that $\eta$ and $\eta_z$ would reach slope $-1/4$ given 
enough time and mass range. Calculation has to be stopped when 
$x$ reaches $\sim 10^8$ because later on biggest planetesimals 
start interacting with bodies of similar size in the shear-dominated 
regime. Such a possibility was not explicitly considered in the 
present study (but it can be easily dealt with using analytical 
apparatus developed in this work).

\begin{figure}
\epsscale{2.0}
\plottwo{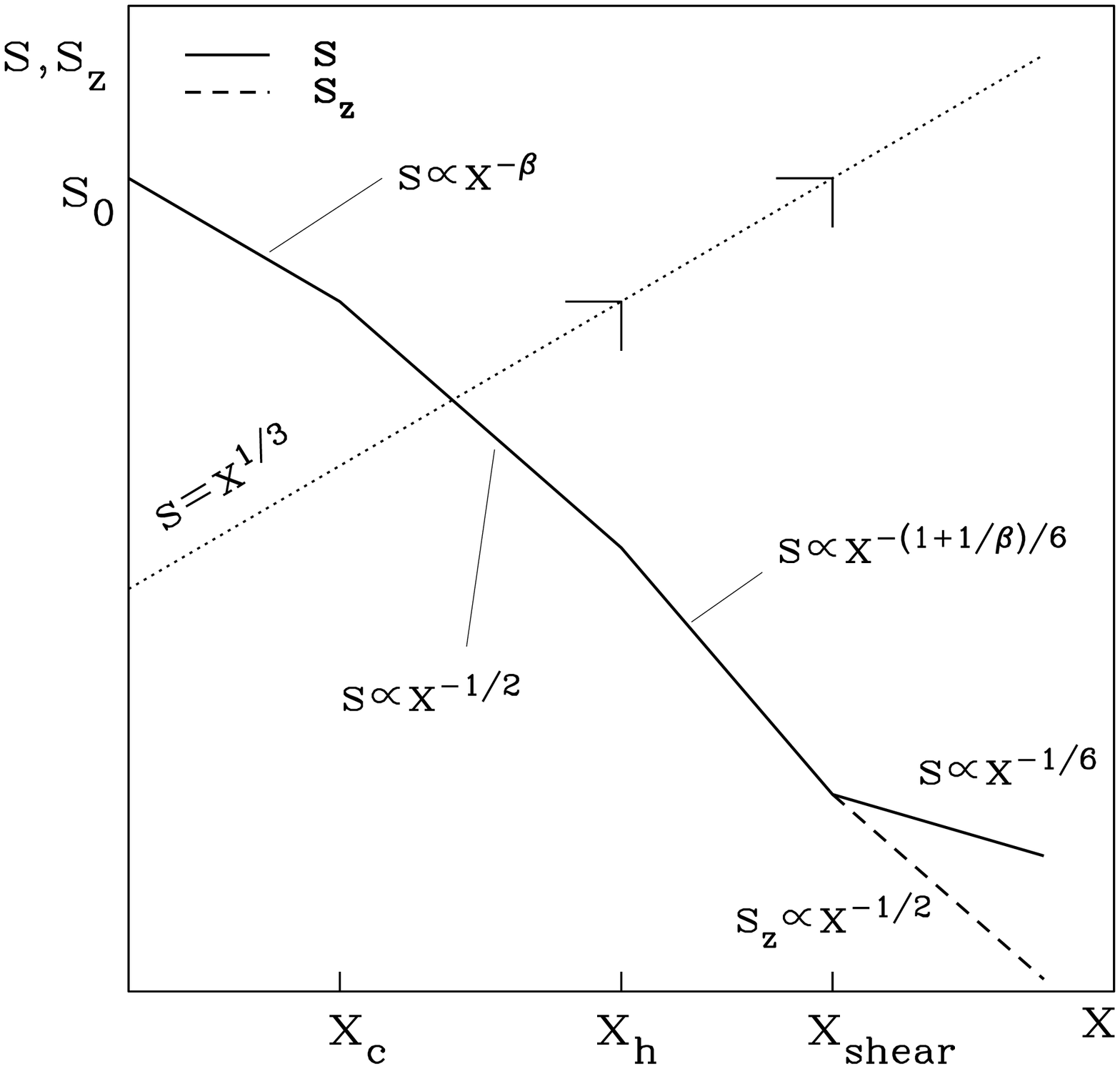}{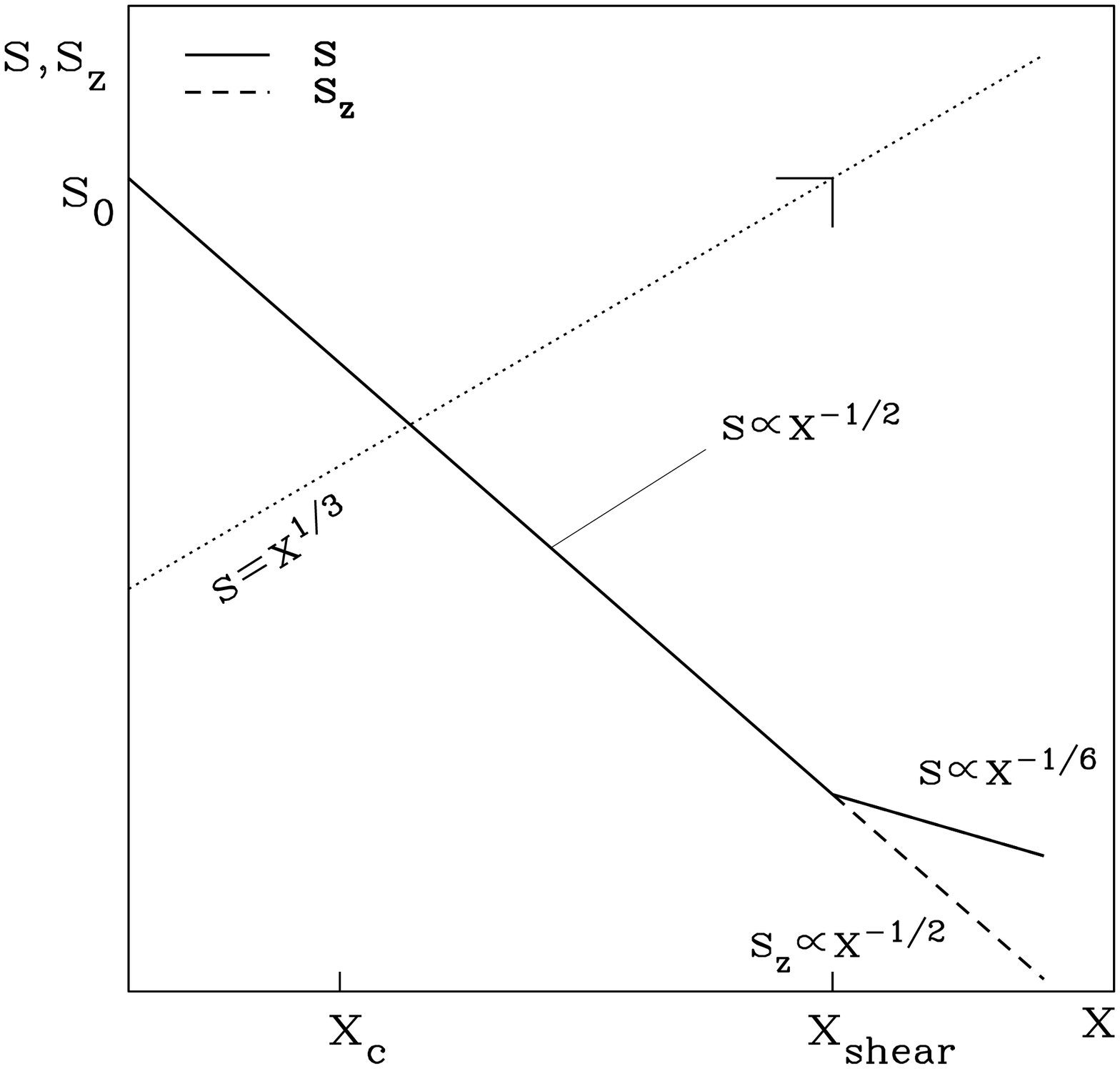}
\caption{
(a) Same as Figure \ref{fig:shallow} but for the Case 1 of steep 
mass spectrum ($3<\alpha<4$). Note that $1/4<\beta<1/2$ 
[see equation (\ref{eq:constr})]. (b) Same as Figure \ref{fig:shallow} 
but for steep mass spectrum with $\alpha>4$. 
\label{fig:less4}}
\end{figure}

Results for biggest runaway bodies ($x\gtrsim x_{shear}$)  
are in nice concordance with analytical predictions given by 
equations (\ref{eq:vel_spec_tailz7}). In the range 
$x_c\lesssim x \lesssim s_{shear}$ velocities of the 
runaway bodies clearly have
not yet converged to a steady state but are evolving in the right 
direction. The range of masses where velocity spectrum should
exhibit a slope $-1/2$ [$x_c\lesssim x\lesssim x_h$, see 
(\ref{eq:vel_spec_tail3})] is so narrow\footnote{
Position of $x_h$ in Figure \ref{fig:num_2.3} is determined by
equation (\ref{eq:x1def}) which assumes that between 
$x_k$ and $x_c$ velocity behaves $\propto x^{-1/4}$. In reality,
as we discussed above, velocity slope in that mass
region is somewhat steeper causing (\ref{eq:x1def}) to
overestimate $x_h$.} for the last (solid) curve 
(because planetesimals with mass $\sim x_c$ are already very close 
to shear-dominated regime) that we do not see this regime 
realized. At the same time, it is conceivable that runaway bodies 
which interact with planetesimals
in the mixed regime ($x_h\lesssim x \lesssim x_{shear}$,
see Appendix \ref{app:subsubsect:tail2}) will finally reach 
their asymptotic velocity state with $\eta=-5/6$ 
and $\eta_z=-7/6$ [see equation (\ref{eq:vel_spec_tailz5})], 
although it will take very long time for them to get there.

Comparison for the case $\alpha=2.7$ (Figure \ref{fig:num_2.7})
is very similar. All major features corresponding to the 
intermediate mass spectrum can be traced in this case as well.
Unfortunately, the comparison with analytical results 
is complicated by the fact that in this case we have to stop 
numerical calculation even earlier than in $\alpha=2.3$ case 
to avoid bringing most massive planetesimals into the 
shear-dominated regime (for this reason mass scale in 
Figure \ref{fig:num_2.7} does not go as far as in Figure 
\ref{fig:num_2.3}). As a result, even a velocity plateau at 
small masses does not have time to fully develop, not 
speaking of runaway bodies in ``mixed'' interaction state. 
Despite this, the overall agreement between the numerical results 
and asymptotic theory of velocity evolution of intermediate mass 
distributions is rather good, especially if we make allowance for 
the short duration of our calculation and small mass range
of planetesimals.

%%%%%%%%%%%%%%%%%%%%%%%%%%%%%%%%%%%%%%%%%%%%%%%%%%%%%%%%%%%
%%%%%%%%%%%%%%%%%%%%%%%%%%%%%%%%%%%%%%%%%%%%%%%%%%%%%%%%%%%

\subsection{Steep mass spectrum.
\label{subsect:steep}}

%%%%%%%%%%%%%%%%%%%%%%%%%%%%%%%%%%%%%%%%%%%%%%%%%%%%%%%%%%%%%%%%%%%

Planetesimal spectra decreasing with mass steeper than $m^{-3}$ are
usually not found in calculations of planetesimal agglomeration. 
Still, this case is quite interesting and we briefly
discuss it here.

\begin{figure}
\epsscale{2.0}
\plottwo{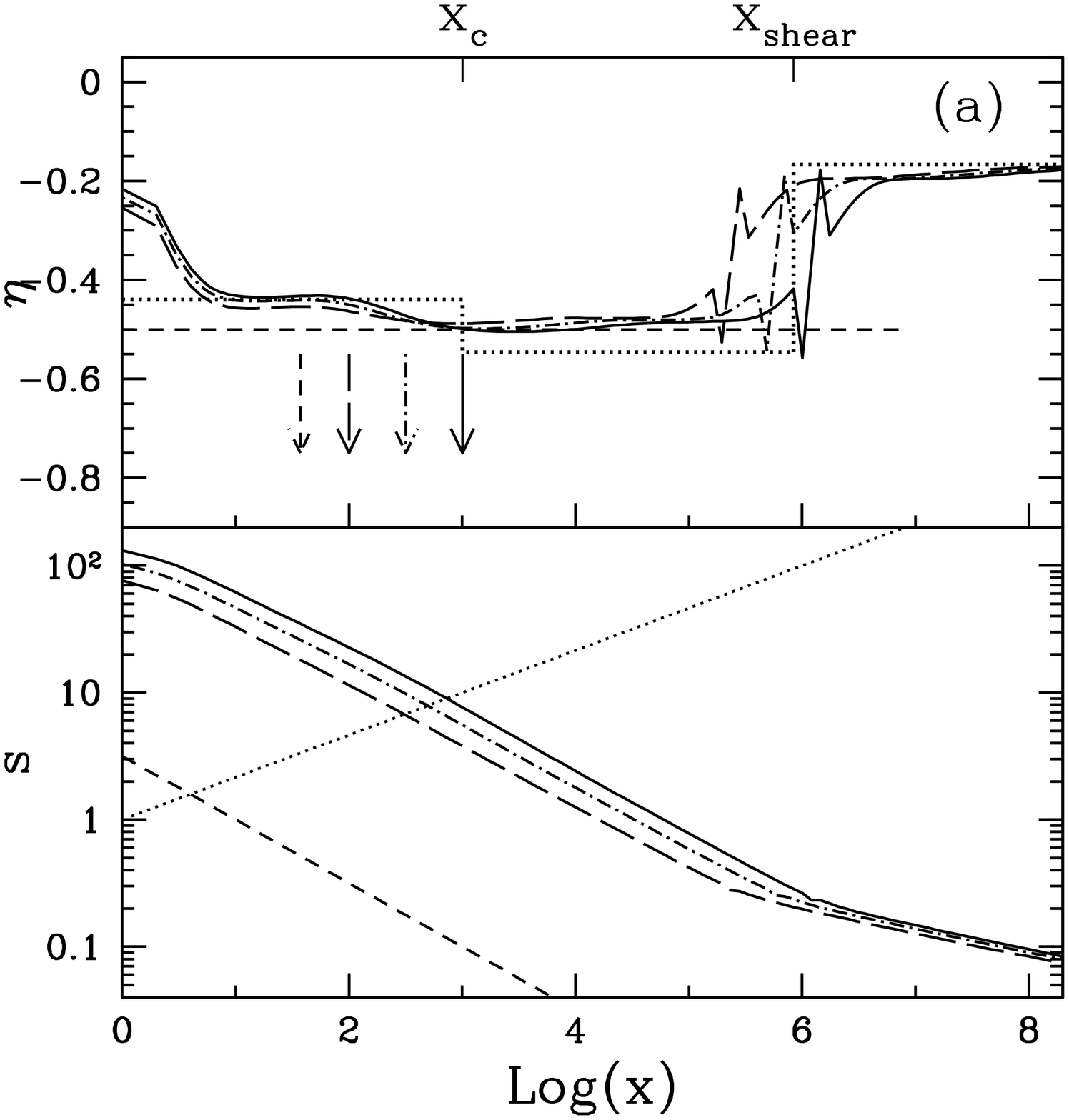}{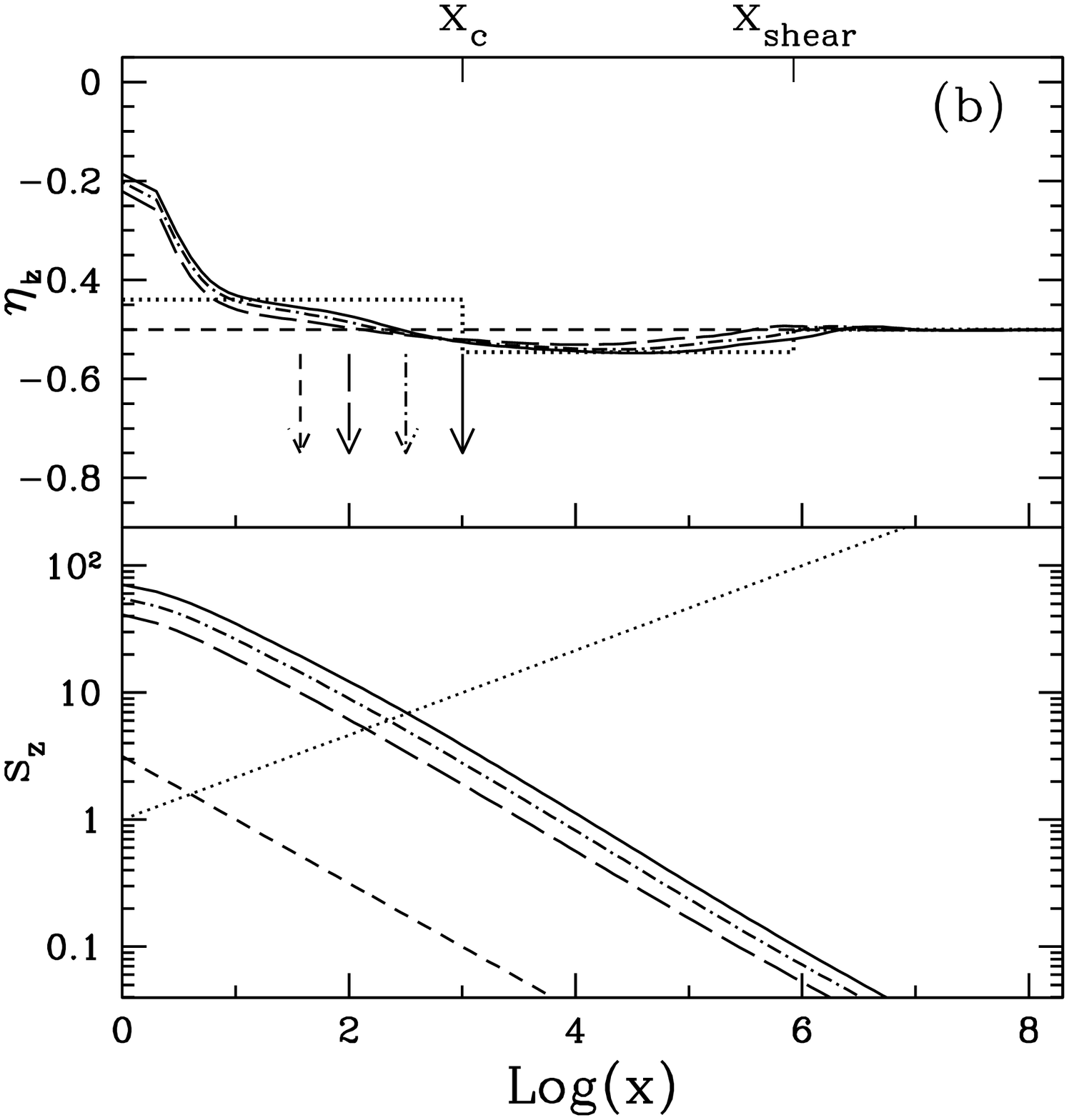}
\caption{
The same as Figure \ref{fig:num_1.5} but for the case of steep 
mass spectrum with  $\alpha=3.2$ (cf. Fig. \ref{fig:less4}a).
\label{fig:num_3.2}}
\end{figure}

As demonstrated in \S 4.1 \& 
\S 4.2 stirring of planetesimals is 
determined solely by biggest bodies with masses $\sim x_c$ 
when $\alpha<3$. This is no longer 
true when the mass spectrum becomes steeper than $x^{-3}$.
Indeed, let us consider the evolution of $s_0$ --- 
velocity dispersion of {\it smallest}
planetesimals having dimensionless mass $x=1$. Using  
equation (\ref{eq:dd_eq}) we find that 
\begin{eqnarray}
&& \frac{\partial s_0^2}{\partial \tau}\approx\frac{1}{s_0^2}
\int\limits_1^{\infty}dx^\star f(x^\star)\frac{x^\star(x+x^\star)}
{1+s^{\star 2}/s_0^2}\nonumber\\
&& \times
\left[\frac{x^\star}{x+x^\star} A_1
-2\frac{1}{1+ 
s^{\star 2}/s_0^2}A_2\right]
\label{eq:dd_eq5}
\end{eqnarray}
Assuming that $s^\star/s_0\to 0$ when $x^\star/x\to \infty$ 
as a result of gravitational friction we find that integral 
in (\ref{eq:dd_eq5}) is dominated by its lower integration limit for 
$\alpha>3$. This means that the velocity of smallest planetesimals is
now mediated by smallest planetesimals themselves, unlike the case of 
$\alpha<3$. We also find that the temporal behavior 
of $s_0$ is given simply by
\begin{eqnarray}
s_0(\tau)\approx \tau^{1/4},~~~~~~~\alpha>3.
\label{eq:vel_spec4}
\end{eqnarray}

We now turn our attention to studying planetesimal velocities in 
the mass range $1\lesssim x\lesssim x_c$. 
We make an a priori assumption that planetesimal velocity spectrum has 
a form of a simple power law (with constant power law index)
\begin{eqnarray}
s(x,\tau)\approx s_0(\tau) x^{-\beta},~~~~~~~x\lesssim x_c,
\label{eq:assume}
\end{eqnarray}
and verify later if it is correct.
In Appendix \ref{app:subsect:steep}
we show that whenever $3<\alpha<4$, 
both heating and friction of planetesimals of a particular 
mass $x$ are determined by bodies of similar mass, $x^\star\sim x$.
This is somewhat unusual in view of our previous 
results when only the extrema
of power-law mass spectrum were important (see 
\S 4.1 and \S 4.2. 
In this case power law index $\beta$ can only be found by
numerically solving equation (\ref{eq:I1I2_cond}); 
there is no simple expression for $\beta$ 
but its value has to satisfy condition
\begin{eqnarray}
\frac{\alpha-2}{4}<\beta<\frac{1}{2}.
\label{eq:constr}
\end{eqnarray}

For even steeper mass distributions, $\alpha>4$, we demonstrate in 
Appendix \ref{app:subsect:steep} that smallest planetesimals 
dominate both stirring and friction not only of 
themselves, but also of bigger bodies. This occurs because
mass spectrum is very steep and high mass planetesimals 
contain too little mass to be of any dynamical importance. 
This is very similar to the velocity evolution of the 
runaway tail in the dispersion-dominated regime described
in \S 4.1.2, and, not surprisingly, we find that
$\beta=1/2$ when  $\alpha>4$. 
Thus, a solution leading to a complete energy equipartition between 
planetesimals ($s\propto x^{-1/2}$) 
is only possible for very steep mass spectra. 
Velocity behavior of runaway bodies in the case
of steep mass spectrum is 
described in detail in Appendix \ref{app:subsect:steep}. 
Theoretical spectra of planetesimal velocities 
are schematically illustrated in Figures 
\ref{fig:less4}a (for $3<\alpha<4$) and 
\ref{fig:less4}b (for $\alpha>4$).

\begin{figure}
\epsscale{2.0}
\plottwo{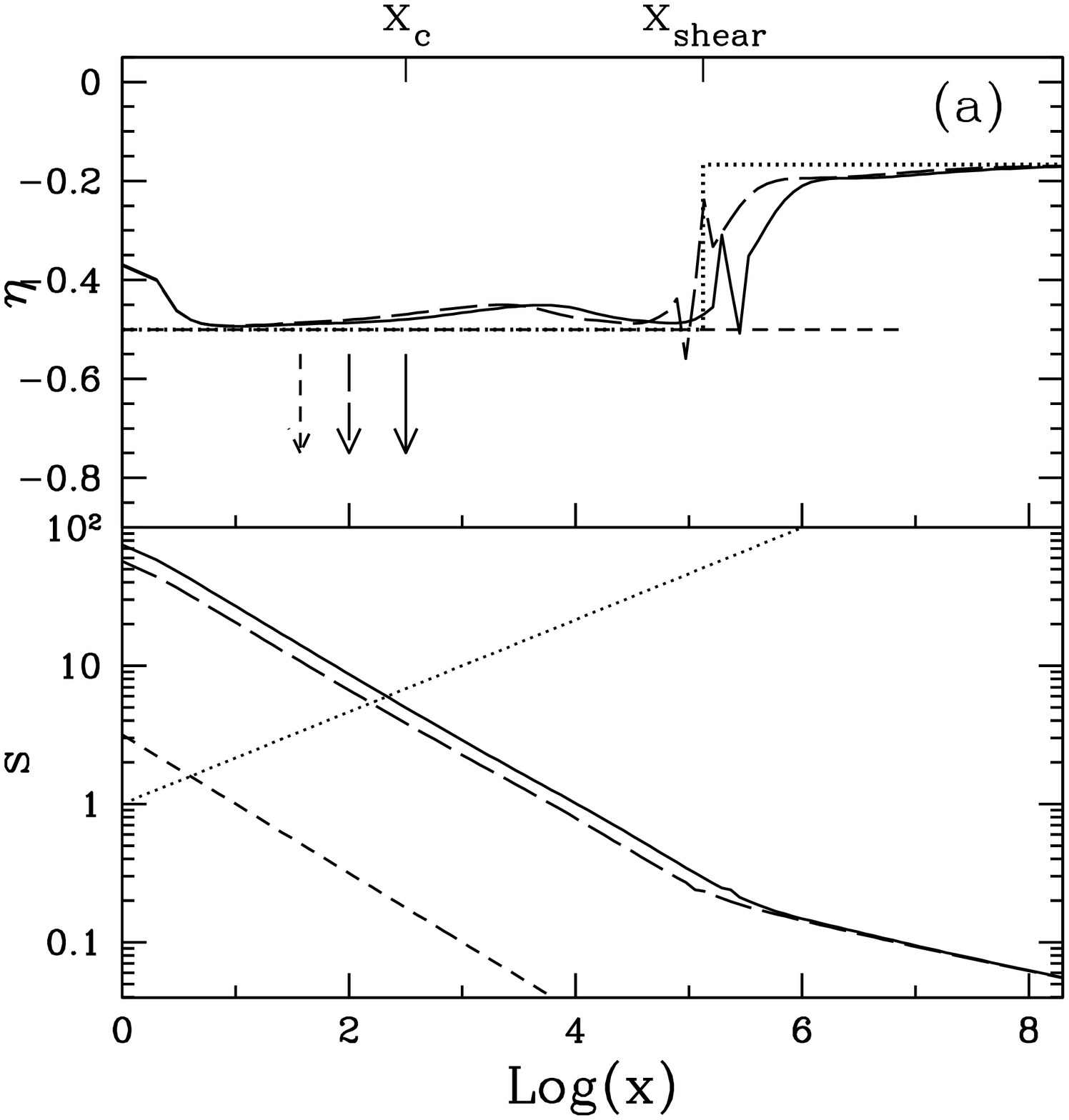}{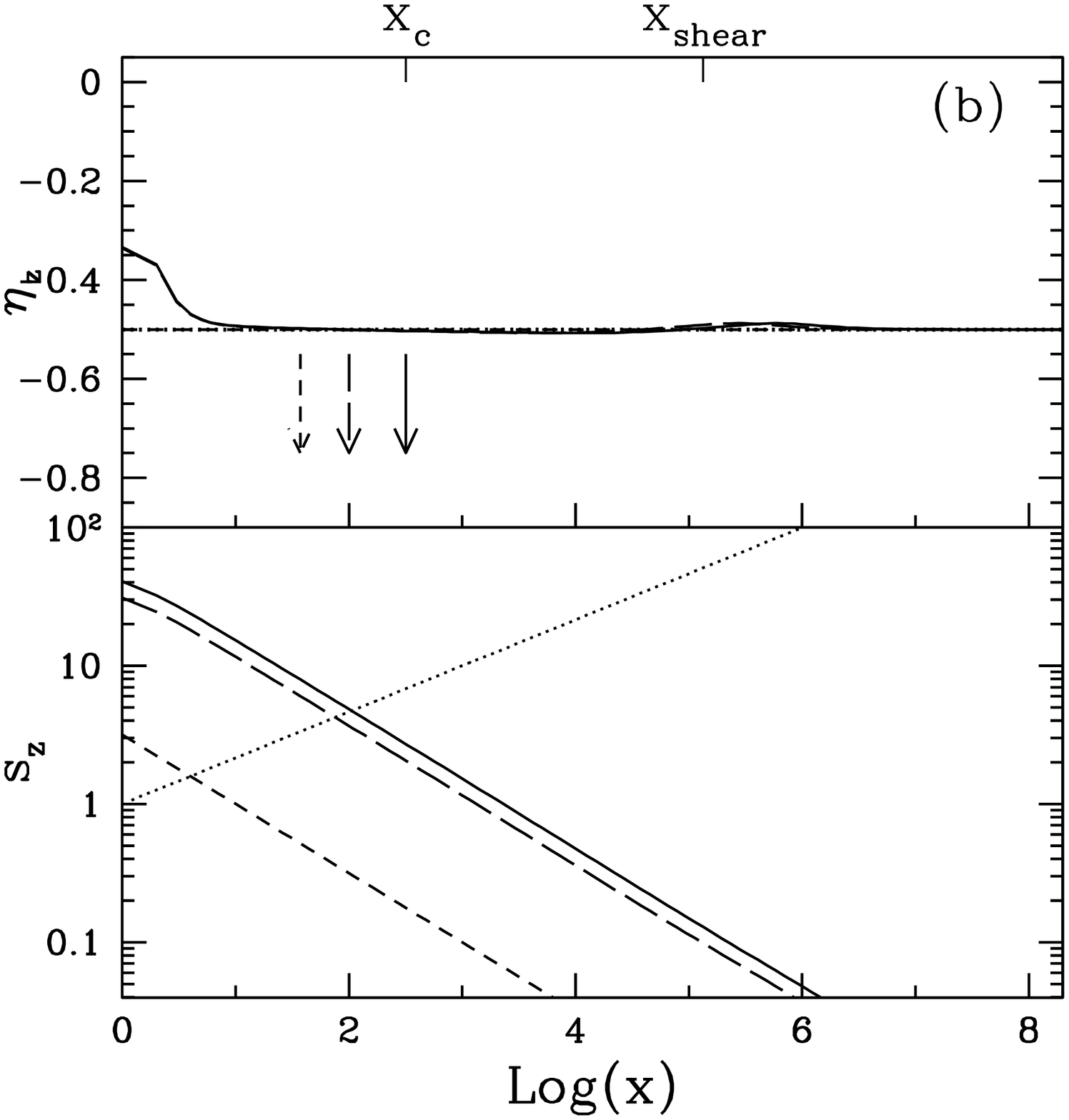}
\caption{
The same as Figure \ref{fig:num_1.5} but for the case of steep
mass spectrum with  $\alpha=4.3$ (cf. Fig. \ref{fig:less4}b).
\label{fig:num_4.3}}
\end{figure}

%Results of this section show that velocities of planetesimals 
%decrease with their mass when size distribution is steep. 
%This inevitably leads at some point to the 
%violation of our assumption of dispersion-dominated scattering
%between all planetesimals. Some of them --- most massive ones
%($\sim x_c$) --- begin to interact in the shear-dominated regime.
%Such a possibility was not considered in present study, 
%but it can be easily dealt with using analytical apparatus 
%advanced in our work.
 
We verify the accuracy of these predictions by numerically 
calculating the
velocity evolution of planetesimal disks with $\alpha=3.2$
and $\alpha=4.3$. In the first case, shown in Figure
\ref{fig:num_3.2}, the slope $\beta$ of planetesimal 
velocity spectrum for $x\lesssim x_c$ is evidently shallower 
than $-1/2$ but steeper than $-(\alpha-2)/4=-0.3$, in 
agreement with constraint (\ref{eq:constr}). 
Analytical prediction
for the behavior of $\eta(x)$ in top panels of Figure 
\ref{fig:num_3.2} is computed using the numerically determined 
value of $\beta$ and equations (\ref{eq:vel_tail_spec8}),
(\ref{eq:vel_tail_spec9}), and (\ref{eq:vel_tail_spec7}). It 
agrees with numerical result (solid curves corresponding to the 
largest $x_c$ displayed) rather well, especially for 
$x\gtrsim x_{shear}$. 

In the case $\alpha=4.3$ (Figure \ref{fig:num_4.3}) the
slope of planetesimal velocity scaling with mass 
is essentially indistinguishable from
$-1/2$ for $x\lesssim x_c$, 
in accordance with our asymptotic prediction. 
Velocity evolution of runaway tail also agrees with the discussion
in Appendix C.2 quite well. 
In both $\alpha=3.2$ and 
$\alpha=4.3$ cases the final velocity curves displayed 
are at the limit where we can still apply our theory: 
$s(x_c)\sim x_c^{1/3}$; this washes out some details 
of velocity spectra of runaway tails 
predicted in Appendix C.2. 
Differences of $\eta$ and $\eta_z$ from their predicted values
at smallest planetesimal masses are caused by boundary effects.
Note that the eccentricity 
dispersion $s$ is virtually independent of time above $x_{shear}$ 
for both $\alpha=3.2$ and $4.3$ (while inclination dispersion $s_z$
slowly grows as $\tau$ increases), which is in complete agreement with 
equation (\ref{eq:vel_tail_spec7}). Finally, numerically computed 
$s_0(\tau)$ follows reasonably well analytical result represented by
formula (\ref{eq:vel_spec4}), which is demonstrated in Figure 
\ref{fig:s_0}d.

%%%%%%%%%%%%%%%%%%%%%%%%%%%%%%%%%%%%%%%%%%%%%%%%%%%%%%%%%%%
%%%%%%%%%%%%%%%%%%%%%%%%%%%%%%%%%%%%%%%%%%%%%%%%%%%%%%%%%%%

\section{Velocity scaling in the presence of dissipative effects.
\label{sect:dissipative}}

Until now we have been assuming that gravity is the only force 
acting on planetesimals. In reality there should be other
factors such as gas drag and inelastic collisions between 
planetesimals which might influence their eccentricities and 
inclinations. Here we briefly comment on the possible changes 
these effects can give rise to.

In the presence of gas drag eccentricity and inclination of a 
particular planetesimal with mass $m$ and physical size $r_p$
tend to decrease with time. Adachi \etal (1976) investigated 
the damping of planetesimal eccentricities and inclinations 
assuming gas drag force proportional to the square of 
planetesimal velocity relative to the gas flow. They came out with 
orbit-averaged prescription for the evolution of random velocity
of planetesimals which can be written at the level 
of accuracy we are pursuing in this study 
(dropping all possible constant factors and assuming that 
eccentricity is of the same order as inclination) as
\begin{eqnarray}
\frac{\partial e^2}{\partial t}\approx -e^2(e+\eta)
\frac{\rho_g \Omega a}{\rho_p r_p},
\label{eq:Adachi}
\end{eqnarray}
where $\rho_g$ is the mass density of nebular gas, $\rho_p$
is the physical density of planetesimal, and $\eta\equiv
(c_s/\Omega a)^2\ll 1$, with $c_s$ being the sound speed of 
nebular gas. In our notation (see \S \ref{sect:velev}) this equation
translates into
\begin{eqnarray}
&& \frac{\partial s^2}{\partial \tau}\approx 
-\zeta s^2 x^{-1/3}\left[s+\eta\left(\frac{M_c}{m_0}
\right)^{1/3}\right],\nonumber\\
&& \zeta\equiv\frac{\Sigma_g}{\Sigma_p}
\frac{\Omega a}{c_s}\left[\frac{m_0 M_c}{(\rho_p a^3)^2}
\right]^{1/3}\nonumber\\
&& \approx 10^{-6} a_{AU}^{-5/2}
\left(\frac{\Sigma_g/\Sigma_p}{100}\right)
\left(\frac{1~{\rm km~s}^{-1}}{c_s}\right)
\left(\frac{m_0}{10^{18}{\rm g}}\right)^{1/3},
\label{eq:gas_drag}
\end{eqnarray}
where $\Sigma_g$ is the surface mass density of the gas disk,
and numerical estimate is made for $\rho_p=3~{\rm g~cm}^{-3}$
and $M_c=M_\odot$. 
For protoplanetary nebula consisting of solar metallicity gas 
one would expect $\Sigma_g/\Sigma_p\approx 50-100$, but during
the late stages of nebula evolution this ratio should be 
greatly reduced by gas removal\footnote{Note that for planet 
formation in some exotic environments such as 
early Universe prior to metal enrichment or metal-poor globular 
clusters this ratio can be much larger than $100$.}.

\begin{figure}
\epsscale{1.0}
\plotone{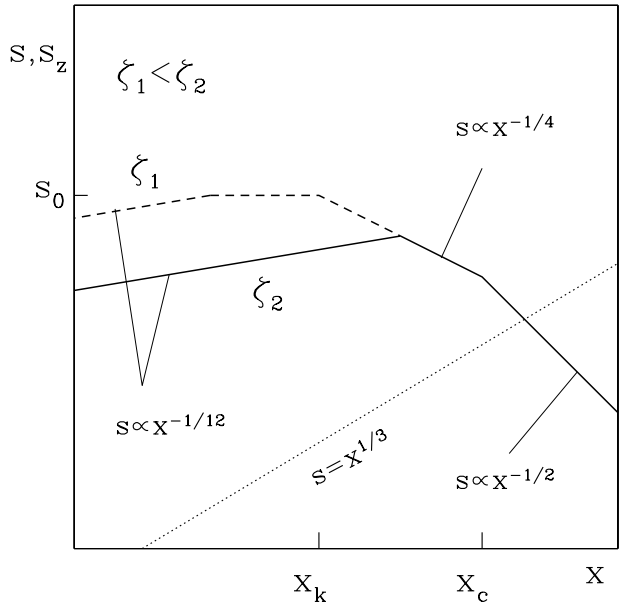}
\caption{
Schematic representation of the effect of gas drag on planetesimal
velocity spectrum in the case $2<\alpha<3$ (cf. Figure 
\ref{fig:interm}). Velocity dependence on
planetesimal mass is displayed for two values of gas drag parameter 
$\zeta$ defined by equation (\ref{eq:gas_drag}): $\zeta_1$
({\it dashed curve}) and $\zeta_2$ ({\it solid curve}), with 
$\zeta_1<\zeta_2$. Velocity spectra overlap at high masses for which
gas drag is no longer important. Inelastic collisions between 
planetesimals would affect their velocity spectrum in a similar manner.
\label{fig:gas_drag}}
\end{figure}

We use the results of \S \ref{sect:scaling} to discuss 
the effect of the gas drag on the planetesimal random velocities 
for mass distributions with $\alpha<3$. We know from 
\S 4.1 and \S 4.2 that 
stirring in this case is dominated by biggest planetesimals
and stirring rate is $\approx M_2(\tau)/s^2$ [see equations
(\ref{eq:dd_eq2}), (\ref{eq:dd_eq3}), and (\ref{eq:dd_eq4})].
Comparing this heating rate with damping due to gas drag
(\ref{eq:gas_drag}) we immediately see that gas drag can become 
important for small planetesimals if planetesimal
velocities are large ($s\gg 1$). In this case dynamical 
excitation by biggest planetesimals is balanced by gad drag. 
Very interestingly, it turns out that such a balance leads to
$s\propto x^{1/12}$ or 
$x^{1/15}$ (depending on
whether $s$ is bigger or smaller than $\eta(M_c/m_0)^{1/3}$), i.e.
smaller planetesimals have smaller random velocities. 
Such a behavior has been previously seen in coagulation simulations
including effects of gas drag on small planetesimals 
(e.g. Wetherill \& Stewart 1993). 
The time dependence of velocities of small planetesimals 
affected by the gas drag should also be different from that given by
(\ref{eq:vel_spec1}) or (\ref{eq:vel_spec2}). In Figure 
\ref{fig:gas_drag} we schematically display the scaling
of $s$ as a function of $x$ in the presence of gas drag for 
$2<\alpha<3$ and different values of $\zeta$.

Inelastic collisions between planetesimals become important for 
their dynamical evolution roughly when relative velocity 
at infinity with which two planetesimals approach each other  
becomes comparable to the escape speed from the biggest of them;
in this case gravitational focusing is inefficient. 
The escape speed   
$v_{esc}\approx 1$ m s$^{-1}r_p/(1~\mbox{km})$ 
(for physical density $\rho=3$ g cm$^{-3}$)
is much larger than the Hill velocity for the same mass
$\Omega R_H\approx 5$ cm s$^{-1}a_{AU}^{-1/2}r_p/(1~\mbox{km})$ 
(for $M_c=M_\odot$), which justifies initially our
previous neglect of inelastic collisions (especially far 
from the Sun where $\Omega R_H$ is small). Later on,
however, continuing dynamical heating in the disk 
would certainly bring velocities of small planetesimals 
above their escape speeds.
When this happens, collisions with bodies of similar size lead to strong
velocity dissipation. This damping should produce similar effect 
on planetesimal random velocities as gas drag does. 
Indeed, kinetic energy losses in physical collisions 
between high velocity planetesimals are roughly proportional 
to their initial kinetic energies, which is similar 
to the behavior of the gas drag losses.

At the same time, velocities of light bodies should still be 
smaller than the
escape velocity from the biggest planetesimals, since otherwise
these planetesimals would not be able to dynamically heat the
disk and it would ``cool'' below their escape velocity. 
Thus, planetesimal velocities cannot 
exceed the escape speed from the surfaces of bodies doing 
most of the stirring, result dating back to a classical 
study by Safronov (1972). This implies that 
gravitational stirring must still be done by the 
highest mass planetesimals (if $\alpha<3$) in accordance 
with what we assumed in the case of gas drag.
As a result, one would again expect 
$s$ to exhibits a power law dependence on mass with positive 
(but small) slope. Exact value of the power law index of 
this dependence is determined by the scaling of energy 
losses with the mass of planetesimals involved in 
collision\footnote{Note that in the crude picture of collision 
energy losses mentioned above $d\sigma_e^2/dt$ is independent of 
planetesimal mass. Then the power law index of velocity spectrum
of small planetesimals is exactly zero. More accurate treatment
of planetesimal collisions might result in some mass dependence
of random velocities.}. 
Effect of the collisions should be most pronounced for small 
planetesimals for which relative encounter velocities can be 
much higher than their $v_{esc}$. Threshold mass at which
highly inelastic collisions become important must constantly increase
in time as planetesimal disk is heated up by massive planetesimals.
One should also remember that  
planetesimal fragmentation in high velocity 
encounters may become important in this regime 
which would significantly complicate 
simple picture we described here. 
 
In summary, dissipative effects are unlikely to affect the
dynamics of massive planetesimals. However, they become important 
for small mass planetesimals and lead to positive correlation 
between random velocities of planetesimals and their masses
(unlike the case of pure gravity studied in \S \ref{sect:scaling}), 
which has been previously observed in
self-consistent coagulation simulations (Wetherill \& Stewart 1993). 
It is thus fairly easy to modify our simple picture developed 
in \S \ref{sect:scaling} to incorporate the effects of gas drag 
and inelastic collisions.

%%%%%%%%%%%%%%%%%%%%%%%%%%%%%%%%%%%%%%%%%%%%%%%%%%%%%%%%%%%
%%%%%%%%%%%%%%%%%%%%%%%%%%%%%%%%%%%%%%%%%%%%%%%%%%%%%%%%%%%

\section{Discussion.
\label{sect:disc}}

Although our study has concentrated on 
a particular case of power-law size distributions (most of the mass is 
locked up in the power-law part of the planetesimal mass spectrum)
its results have a wider range of applicability. The reason for this 
is that the velocity scalings derived in \S \ref{sect:scaling}
depend only on what part of planetesimal spectrum dominates stirring 
and friction and not on the exact shape of mass distribution. 
For example, suppose that 
$f(x,\tau)$ does not have a self-similar 
power law form of the kind studied in \S 4.2 
(intermediate mass spectrum) but its first moment $M_1$ is still 
dominated by the smallest masses (most of the 
mass is locked up in smallest planetesimals, which dominate friction), 
while the second moment $M_2$ is mainly determined by highest masses 
(heating is 
determined by biggest planetesimals $\sim x_c$). Then it is easy to see
from our discussion in \S 4.2.1
that the velocity distribution has a form of a broken power-law 
in mass, changing its slope from $0$ at small masses to $-1/4$ at high masses
[see (\ref{eq:vel_spec2}) and (\ref{eq:vel_spec3})]. 
If, on the contrary, most of the mass is not in smallest bodies but 
in largest ones ($M_1$ and $M_2$ converge near the cutoff of 
planetesimal spectrum, $\sim x_c$), then both friction and stirring are 
dominated by biggest planetesimals and velocity spectrum must 
have zero slope for the entire range of planetesimal masses,
in accord with our consideration of shallow mass distributions 
(\S 4.1.1). Whenever both stirring and cooling 
of planetesimals of mass $m$ are dominated by planetesimals of similar 
mass, we go back to the Case 1 of \S 4.3. Finally,
in all cases when stirring and friction 
are produced by planetesimals much smaller 
than those under consideration, velocity profile scales like $m^{-1/2}$
(energy equipartition, see Case 2 in \S 4.3).
Similar rules can be derived on the basis of the results of
\S \ref{sect:scaling} for more complicated mass spectra, and this 
makes our approach very versatile. 

One of the interesting features found in our study of
purely gravitational scattering (see \S \ref{sect:scaling}) 
is the development of
a plateau at the low mass end of planetesimals velocity distribution 
for shallow and intermediate mass spectra ($\alpha<3$). These mass
spectra are among the most often realized in the coagulation 
simulations which makes this prediction quite important for interpretation
of numerical results. One should be cautious enough not to confuse 
these plateaus with the manifestations 
of dissipative processes such as gas drag
or inelastic collisions which tend to endow velocity distribution
with only very weak (positive) dependence on mass, 
see \S \ref{sect:dissipative}.
%Indeed, when a large scale simulations 
%incorporating a lot of physical phenomena endows velocity distribution  
%with a zero-slope part at low masses there is always a strong 
%temptation to ascribe the emergence of such a plateau to the 
%action of gas drag or energy damping in inelastic collisions 
%between planetesimals. Our study demonstrates that although 
%such a conclusion might be correct to some extent, plateau in the 
%velocity profile appears even without any dissipative mechanisms operating. 

Another observation which can be made based on our results is that the
bulk of planetesimals (power law part where most of the mass is locked up) 
rarely reaches a complete equipartition in energy. Equipartition is 
realized when $\sigma_{e,i}\propto m^{-1/2}$, or in our notation
$s, s_z\propto x^{-1/2}$. Our calculations (\S 4.3) 
show that when $x\lesssim x_c$ 
such a velocity spectrum is only possible for very steep mass distributions,
$\alpha>4$, when both dispersion-dominated stirring and friction  
are due to the smallest planetesimals with masses  
$m\sim m_0$ ($x\sim 1$). Mass distributions that steep are not often encountered 
in self-consistent coagulation or N-body simulations (Wetherill \& Stewart 1993;
Kenyon 2002; Kokubo \& Ida 1996, 2000) and yet the equipartition 
argument is very often used to draw important conclusions about the bulk of 
planetesimal population (see below).

At the same time, random energy equipartition 
is ubiquitous for runaway bodies in the high mass tail ($x\ga x_c$) if 
they still interact with the rest of planetesimals in the dispersion-dominated
regime. Since, by our assumption, runaway tail contains very little mass,
this equipartition is very similar to the case of $\alpha>4$, because in 
both situations stirring as well as friction are driven by planetesimals 
much lighter than the bodies under consideration. 
It is also worth mentioning that this conclusion changes when 
runaway bodies start interacting with small planetesimals in the 
shear-dominated regime (see e.g. \S 4.1.2).

Although we did not follow the evolution of the planetesimal mass spectrum 
self-consistently we can already constrain some of the theories explaining
mass distributions in coagulation cascades. For example, 
N-body simulations of gravitational agglomeration of $10^3-10^4$ massive 
planetesimals by Kokubo \& Ida (1996, 2000) produce roughly power law 
mass spectrum with an index of $\approx -2.5$ within some 
range of masses. To explain this result Makino \etal (1998) explored 
planetesimal growth in the dispersion-dominated 
regime with strong gravitational focusing. They postulated 
epicyclic energy to be in equipartition and found that 
planetesimal mass distribution 
has a power-law form with a slope of $-8/3\approx -2.7$. 
%Planetesimals in a disk with such a mass spectrum would grow mainly by collisions
%with bodies of similar mass. 
However, using the results of \S 4.2.1 one can 
immediately pinpoint the contradiction with the
 basic assumptions which lead to this conclusion. 
Indeed, slope of $-8/3$ corresponds to the intermediate mass spectrum (see \S 
4.2) which can never reach energy 
equipartition. Moreover, this mass spectrum cannot even have a 
power-law velocity distribution with a continuous slope 
necessary for a theory of Makino \etal (1998) to work. As a result, the 
explanation provided by these 
authors for the mass spectrum found in N-body simulations 
cannot be self-consistent. More work in this direction has to be done.

%Unfortunately, it is not easy to compare our analytical results for 
%purely gravitational scattering with similar works by other authors. 
Although theory developed in \S \ref{sect:scaling} 
is in good agreement with our numerical calculations, one should bear
in mind that our analytical results are asymptotic by construction, 
i.e. they are most accurate when planetesimal disk has evolved for
a long time and when planetesimal size distribution spans a wide range 
in mass (this can be easily seen by comparing
Figures \ref{fig:num_2.3} and \ref{fig:num_2.7}). Real 
protoplanetary disks during their evolution might not enjoy the 
luxury of having such conditions been fully realized; 
this can complicate quantitative applications of our 
results to real systems. We believe 
however that this does not devalue our analytical results 
because they provide basic understanding of processes involved 
in planet formation, allow one to clarify important trends 
seen in simulations, and enable quick and efficient classification
of  possible evolutionary outcomes for a wide range of 
parameters of protoplanetary disks.
 
Several previous 
studies have concentrated on exploring velocity equilibria 
in systems with nonevolving mass spectra in which inelastic 
collisions between planetesimals acted as damping mechanism 
necessary to balance gravitational stirring (Kaula 1979; Hornung, 
Pellat, \& Barge 1985; Stewart \& Wetherill 1988). 
We have outlined a qualitative picture 
of the effects of dissipative processes in \S \ref{sect:dissipative} 
and it agrees rather well with these previous investigations. 
N-body calculations and ``particle in a box'' coagulation 
simulations represent another class of studies in which 
velocity distributions have been routinely computed. As 
we already commented in \S \ref{sect:intro} 
these calculations usually absorb a lot of diverse 
physical phenomena which makes them not easy targets 
for comparisons. Still, there are several generic 
features almost all simulations agree upon 
such as (1) roughly power-law mass spectra typically 
with slopes $\alpha<3$ within some mass range, 
(2) velocity dispersion is constant or slowly increases 
with mass for small planetesimals, and (3) it decreases
with mass for large bodies. This general picture is in good 
agreement with our analytical predictions.

%%%%%%%%%%%%%%%%%%%%%%%%%%%%%%%%%%%%%%%%%%%%%%%%%%%%%%%%%%%
%%%%%%%%%%%%%%%%%%%%%%%%%%%%%%%%%%%%%%%%%%%%%%%%%%%%%%%%%%%

\section{Conclusions.
\label{sect:concl}}

We have carried out an exhaustive census of the dynamical properties
of planetesimal disks characterized by a variety of mass 
distributions. Here we briefly summarize our results for the case 
velocity evolution driven by gravitational 
scattering of planetesimals. 

Whenever planetesimal mass distribution in a disk has a power law
slope shallower than $-2$, we find that planetesimal velocity is
constant up to the upper mass cutoff. For slopes between $-2$
and $-3$ velocity dispersions are constant at small masses but 
switch to $m^{-1/4}$ dependence above some intermediate mass. 
As a result, mass spectrum exhibits a pronounced ``knee'' at this 
mass. Distributions with 
slopes between $-3$ and $-4$ have purely power-law velocity spectra with
slope $-\beta$ satisfying $1/4<\beta<1/2$, which can be determined 
numerically using equation (\ref{eq:I1I2_cond}). Finally, planetesimals 
in disks characterized by mass spectra steeper than $m^{-4}$ are in 
energy equipartition, i.e. velocity dispersions scale as $m^{-1/2}$.
Such a difference in behaviors is caused by the fact that gravitational 
stirring and dynamical friction receive major contributions 
from different parts of planetesimal mass spectra in different cases.

We also consider a possibility that mass distribution has a tail of
massive ``runaway'' bodies sticking out beyond the exponential cutoff
of the power-law mass spectrum. 
We have found that the low mass end of the tail experiencing 
dispersion-dominated scattering by all planetesimals exhibits complete 
energy equipartition: $\sigma_{e,i}\propto m^{-1/2}$. Most massive 
runaway bodies interact with all planetesimals in the shear-dominated 
regime, which leads to highly anisotropic velocity distributions of these 
bodies: $\sigma_e\propto m^{-1/6}$ while $\sigma_i\propto m^{-1/2}$. 
%There is also an intermediate range of masses of runaway bodies 
%where they interact with small planetesimals in the 
%dispersion-dominated regime and with big ones in shear-dominated 
%regime. Velocity scalings in this case depend on the velocity 
%behavior of small-mass planetesimals and  differ for different 
%mass spectra. 
Time dependence of velocities 
is determined mainly by the evolution of the cutoff of the power
law mass spectrum $m_0 x_c(\tau)$. These asymptotic results derived 
using analytical means are in good agreement with numerical calculations
of planetesimal dynamical evolution presented in \S \ref{sect:scaling}.
They can be easily generalized to cover more complicated planetesimal 
mass distributions.

Finally, we investigate qualitatively the impact
of damping processes such as gas drag or inelastic collisions on the
planetesimal disk dynamics. We find that manifestations of these
effects are only important for small mass planetesimals; they exhibit 
themselves in the form of gentle increase of planetesimal random 
velocities with mass which has been previously observed in 
``particle in a box'' coagulation simulations (Wetherill \& Stewart 1993). 

%Our current study has concentrated only on exploring the effects of gravity 
%on the dynamics of the planetesimal disk. We 
%neglected additional complications such as presence of gas,
%physical collisions between bodies, etc. This precludes us from comparing
%our analytical predictions with the results of numerical simulations
%which usually try to model a variety of physical processes.
%However, our own numerical calculations find good agreement with 
%asymptotic theory developed in \S \ref{sect:scaling} using analytic means.

Future work in this direction should target the 
self-consistent coupling of the evolution of planetesimal velocities to
the evolution of mass spectrum of planetesimals due to their coagulation.
Successful solution of this problem would greatly contribute to our 
understanding of how terrestrial planets grow out of swarms of  
planetesimals in protoplanetary disks.

\acknowledgements

I am indebted to Peter Goldreich for inspiration and 
valuable advice I have been receiving 
during all stages of this work. I am also grateful to him and 
Re'em Sari for careful reading of this manuscript and making a 
lot of useful suggestions. Financial support of this work by 
W. M. Keck Foundation is thankfully acknowledged.

\appendix

%%%%%%%%%%%%%%%%%%%%%%%%%%%%%%%%%%%%%%%%%%%%%%%%%%%%%%%%%%%%%%%%%%%

\section{Numerical procedure.
\label{app:numerical_procedure}}

Evolution of planetesimal eccentricities and inclinations 
is performed by numerically evolving in time the 
full system (\ref{eq:homog_heating}) with 
scattering coefficients smoothly interpolated between shear- and 
dispersion-dominated regimes. Planetesimals of different 
masses are distributed in logarithmically 
spaced mass bins, with bodies in each bin being more massive
than planetesimals in previous bin by $1.2$. 
Using numerical orbit integrations 
we have determined the values of constant coefficients in
(\ref{eq:cold_coeffs}) relevant for the shear-dominated scattering
to be 
\begin{eqnarray}
B_1\approx 11.4,~~B_2\approx 4.0,~~D_1\approx 11.4,~~D_2\approx 4.2.
\label{eq:shear_coeffs}
\end{eqnarray}
Scattering coefficients in the dispersion-dominated regime
are taken from Stewart \& Ida (2000) who
derived analytical expressions for viscous stirring and dynamical  
friction rates in the two-body approximation. We translate their 
results for our gravitational stirring and friction functions 
$H_{1,2}$ and $K_{1,2}$. This provides us with the 
dependence of coefficients $A_{1,2}$ and $C_{1,2}$ in 
(\ref{eq:h1h2}) on the inclination to eccentricity ratio 
$\sigma_i/\sigma_e$. Constants $a_{1,2}$ in (\ref{eq:Coulomb}) 
are fixed to be $a_1=1.7, a_2=1$.

We assume that mass scale evolves as 
$x_c(\tau)=F+(\varepsilon \tau)^{\chi}$, where $F\approx 40,~
\varepsilon=10^{-4}$, and $\chi=4$ in the case of shallow mass 
spectrum with $\alpha=1.5$ and $\chi=1.5$ in all other cases. Small
parameter $\varepsilon$ is introduced to mimic the difference between
the timescale of dynamical evolution of planetesimal disk and
the timescale of planetesimal growth (which is usually much longer).
Besides this, such a form of $x_c(\tau)$ is chosen arbitrarily. 
Runaway tail is assumed to have a power-law form with the same slope 
$\alpha$ but much smaller normalization than planetesimal 
mass distribution.
Tail always extends $8$ orders of magnitudes beyond $x_c$ to allow 
interesting velocity regimes to fully develop. Initial distribution
of planetesimal eccentricities and inclinations is set to
$s(x,0)=s_z(x,0)=3x^{-1/2}$.

%%%%%%%%%%%%%%%%%%%%%%%%%%%%%%%%%%%%%%%%%%%%%%%%%%%%%%%%%%%%%%%%%%%

\section{Velocity evolution in a ``mixed'' state for 
intermediate mass spectrum.
\label{app:subsubsect:tail2}}

In the case of intermediate mass spectrum runaway bodies with 
masses $x$ between $x_h$ [defined in (\ref{eq:x1def})] 
and $x_{shear}$ interact with small planetesimals 
[less massive than $x_s(x)$ defined by (\ref{eq:xsx_def})] in 
the dispersion-dominated regime, 
but with large ones (heavier than $x_s(x)$) 
in the shear-dominated regime.
As a result, the equations for $s$ and $s_z$
are now hybrid versions of (\ref{eq:dd_eq_tail}) and (\ref{eq:sd_eq}):
\begin{eqnarray}
&& \frac{\partial s^2}{\partial \tau}=A_1 
\int\limits_1^{x_s(x)}dx^\star \frac{x^{\star 2} f(x^\star)}
{s^{\star 2}}-2A_2 s^2 x
\int\limits_1^{x_s(x)}dx^\star \frac{x^\star f(x^\star)}
{s^{\star 4}}+
\frac{C_1}{x^{2/3}}
\int\limits_{x_s(x)}^{\infty}dx^\star x^{\star 2}f(x^\star)
-
\frac{2C_2s^2}{x^{1/3}}
\int\limits_{x_s(x)}^{\infty}dx^\star x^\star f(x^\star),
\label{eq:hybrid}\\
&& \frac{\partial s_z^2}{\partial \tau}=B_1 
\int\limits_1^{x_s(x)}dx^\star \frac{x^{\star 2} f(x^\star)}
{s^{\star 2}}-2B_2 s_z^2 x
\int\limits_1^{x_s(x)}dx^\star \frac{x^\star f(x^\star)}
{s^{\star 4}}+
\frac{D_1}{x^{4/3}}
\int\limits_{x_s(x)}^{\infty}dx^\star x^{\star 2}
s^{\star 2} f(x^\star)
-
\frac{2 D_2 s_z^2}{x^{1/3}}
\int\limits_{x_s(x)}^{\infty}dx^\star x^\star f(x^\star).
\label{eq:hybridz}
\end{eqnarray}
Following the approach taken in \S 4.1.2 
we neglect l.h.s. 
of both equations and use (\ref{eq:vel_spec2}) 
and (\ref{eq:vel_spec3}) to evaluate integrals in evolution 
equations. After careful comparison of different contributions
in the r.h.s. we find that
\begin{eqnarray}
&& C_1\frac{M_2}{x^{2/3}}\approx 2A_2 M_1\frac{s^2 x}{s_0^4},
\label{eq:hybrid1}
\\
&& B_1\frac{x_s^{7/2-\alpha}}{s_0^2x_k^{1/2}}+
D_1\frac{s_0^2 x_k^{1/2}}{x^{4/3}}\int\limits_{\sim x_s}^\infty
dx^\star (x^\star)^{3/2}f(x^\star)
\approx 2B_2 M_1\frac{s_z^2 x}{s_0^4}.
\label{eq:hybridz1}
\end{eqnarray}
Equation (\ref{eq:hybrid1}) implies that eccentricity stirring of 
runaway  bodies in the mass range $x_h\ll x\ll x_{shear}$ 
is done mainly  
by planetesimals with masses $\sim x_c$, for which the  
interaction proceeds in the {\it shear}-dominated regime.
This heating is balanced by dynamical friction due to the smallest 
planetesimals, lighter than $x_k$, which interact with 
these massive bodies in the  
{\it dispersion}-dominated regime.

Vertical heating is somewhat different. Gravitational friction is 
again dominated by smallest planetesimals which are in the 
dispersion-dominated mode. Stirring, however, depends on
shape of the mass distribution. 
If $\alpha< 5/2$, second term in the l.h.s. of 
(\ref{eq:hybridz1}) --- shear-dominated heating 
by planetesimals with masses between $\sim x_s$ and $\sim x_c$ 
--- dominates\footnote{For $\alpha<5/2$ integral in 
(\ref{eq:hybridz1}) converges at the upper end of its range,
near $x_c$. When $\alpha=5/2$ this integral is contributed 
roughly equally by equal logarithmic intervals in mass between
$\sim x_s$ and $\sim x_c$; then the second term in the l.h.s. 
of (\ref{eq:hybridz1}) 
dominates over the first one by $\sim\ln(x_c/x_s)$.} 
over the first term which represents the effect 
of dispersion-dominated heating by planetesimals with masses 
$\sim x_s$. In the opposite case, when $\alpha>5/2$ both terms 
in the l.h.s. of (\ref{eq:hybridz1}) 
produce roughly equal contributions.
Using (\ref{eq:hybrid1}) \& (\ref{eq:hybridz1}) 
one can easily derive expressions 
(\ref{eq:vel_spec_tail4})-(\ref{eq:vel_spec_tailz5}).

%%%%%%%%%%%%%%%%%%%%%%%%%%%%%%%%%%%%%%%%%%%%%%%%%%%%%%%%%%%
%%%%%%%%%%%%%%%%%%%%%%%%%%%%%%%%%%%%%%%%%%%%%%%%%%%%%%%%%%%

\section{Details of velocity evolution for the case of steep mass spectrum.
\label{app:subsect:steep}}

Substituting our guess 
(\ref{eq:assume})
into (\ref{eq:dd_eq}) and going
through all possibilities appropriate for 
$\alpha>3$
we find that self-consistent solutions of power-law type
can exist in only two cases:
\begin{itemize}

\item Case 1: ~~~$3-\alpha+2\beta>0$~~~ \&~~~ $2-\alpha+4\beta>0$.

\item Case 2: ~~~$3-\alpha+2\beta<0$~~~ \&~~~ $2-\alpha+4\beta<0$,

\end{itemize}
We now consider separately these two possibilities.

%%%%%%%%%%%%%%%%%%%%%%%%%%%%%%%%%%%%%%%%%%%%%%%%%%%%%%%%%%%%%%%%%%%

\subsection{Planetesimal velocities.
\label{app:subsubsect:powlaw3}}

%%%%%%%%%%%%%%%%%%%%%%%%%%%%%%%%%%%%%%%%%%%%%%%%%%%%%%%%%%%%%%%%%%%

{\bf Case 1.}

%%%%%%%%%%%%%%%%%%%%%%%%%%%%%%%%%%%%%%%%%%%%%%%%%%%%%%%%%%%%%%%%%%%

In this case one can easily see that integrals in the r.h.s. 
of (\ref{eq:dd_eq}) are mostly contributed by $x^\star\sim x$.
This allows us to rewrite (\ref{eq:dd_eq}) in the following form:
\begin{eqnarray}
x^{-2\beta}\frac{\partial s_0^2}{\partial \tau}=
\frac{x^{3-\alpha+2\beta}}{s_0^2}
\int\limits_0^{\infty}dt \frac{t^{1-\alpha}(1+t)}
{1+t^{-2\beta}}\left[\frac{t}{1+t} A_1
-2\frac{1}{1+t^{-2\beta}}A_2\right].
\label{eq:dd_eq6}
\end{eqnarray}
In arriving at this expression we have used the fact that 
$f(x,\tau)=x^{-\alpha}$ for $\alpha>3$ and $1\ll x\ll x_c$, 
and extended the integration range over $t\equiv x^\star/x$
from $0$ to $\infty$ (because only local region $t\sim 1$
matters). From this equation we can readily see that the l.h.s. 
of (\ref{eq:dd_eq6}) can 
be neglected for $x\gg 1$. As a result, a self-consistent power-law
solution for $s$ exists only if integral in (\ref{eq:dd_eq6}) is
equal to zero. This leads to the following constraint on the 
required value of $\beta$:
\begin{eqnarray}
\frac{I_1(\alpha,\beta)}{I_2(\alpha,\beta)}=\frac{A_1}{2A_2},~~~
I_1(\alpha,\beta)\equiv\int\limits_0^\infty dt
\frac{t^{1-\alpha}(1+t)}{(1+t^{-2\beta})^2},~~~
I_2(\alpha,\beta)\equiv\int\limits_0^\infty dt
\frac{t^{2-\alpha}}{1+t^{-2\beta}}.
\label{eq:I1I2_cond}
\end{eqnarray} 
Solving this equation for given $\alpha$ and $A_1/A_2$ one
can find the slope of the mass spectrum $\beta(\alpha,A_1/A_2)$.
At a first sight it is not at all clear that equation 
(\ref{eq:I1I2_cond}) should in general possess a solution for 
$\beta$. However, closer look at the problem reveals 
some interesting patterns.

First of all, it follows  from (\ref{eq:property})
that r.h.s. of (\ref{eq:I1I2_cond}) has to be bigger than $1$.  
Second, in the case of energy equipartition  
$I_1(\alpha,1/2)/I_2(\alpha,1/2)=1$. Third, 
$I_1(\alpha,\beta)\to\infty$ as $\beta\to (\alpha-2)/4$, while
$I_2$ stays finite. These observations prove that 
(\ref{eq:I1I2_cond}) {\it always} has a solution 
for $\beta$ satisfying the 
constraint posed by equation (\ref{eq:constr}).
Combining restriction (\ref{eq:constr})
with initial constraints of Case 1
we find that Case 1 is only possible for mass spectra 
with power law slope $3<\alpha<4$.

In fact, one can do things even better by considering separately
eccentricity and inclination scalings with mass, i.e. using both equations 
(\ref{eq:homog_heating_integr}). 
Assuming that the ratio of inclination to eccentricity 
$s_z/s$ is still constant for $x\lesssim x_c$, 
one would obtain in addition to (\ref{eq:I1I2_cond}) 
another equation
dictated by the inclination evolution. It would be identical to   
(\ref{eq:I1I2_cond}), but with $A_{1,2}$ replaced by $B_{1,2}$.
Since both $A_{1,2}$ and $B_{1,2}$ depend only on $s_z/s$ (see \S 
\ref{sect:velev}) one would have two equations for two unknowns: $\beta$ and
$s_z/s$. Solving them one can uniquely fix both the power 
law index of planetesimal velocity dependence on mass and the 
ratio of inclination to eccentricity of planetesimals (which is left 
undetermined in our simplified analysis of \S \ref{sect:scaling}).

%%%%%%%%%%%%%%%%%%%%%%%%%%%%%%%%%%%%%%%%%%%%%%%%%%%%%%%%%%%%%%%%%%%

{\bf Case 2.}

%%%%%%%%%%%%%%%%%%%%%%%%%%%%%%%%%%%%%%%%%%%%%%%%%%%%%%%%%%%%%%%%%%%

Restrictions imposed on $\alpha$ and $\beta$ by the conditions of 
Case 2 imply that all integrals in the r.h.s. of (\ref{eq:dd_eq})
are dominated by the lower end of their integration range, i.e.
by masses $\sim 1$. 
Neglecting time derivative in (\ref{eq:dd_eq}) and balancing 
contributions in the r.h.s. we find that $\beta=1/2$ and
\begin{eqnarray}
s(x,\tau)\approx s_0(\tau)x^{-1/2},~~~~~x\lesssim x_c.
\label{eq:vel_spec5}
\end{eqnarray}
This solution demonstrates that for $x\gg 1$ our omission
of the l.h.s. of (\ref{eq:dd_eq}) is justified for arbitrary behavior
of $x_c(\tau)$. 
It also imposes important restriction on the mass spectra for which 
this scaling can be realized: $\alpha$ mus be bigger than $4$. 
This means that we have found 
solutions for all possible positive values of $\alpha$ and 
our study is complete in this sense.

%%%%%%%%%%%%%%%%%%%%%%%%%%%%%%%%%%%%%%%%%%%%%%%%%%%%%%%%%%%%%%%%%%%
%%%%%%%%%%%%%%%%%%%%%%%%%%%%%%%%%%%%%%%%%%%%%%%%%%%%%%%%%%%%%%%%%%%

\subsection{Velocities of runaway bodies.
\label{app:subsubsect:tail3}}

%%%%%%%%%%%%%%%%%%%%%%%%%%%%%%%%%%%%%%%%%%%%%%%%%%%%%%%%%%%%%%%%%%%

{\bf Case 1 ($3<\alpha<4$).}

%%%%%%%%%%%%%%%%%%%%%%%%%%%%%%%%%%%%%%%%%%%%%%%%%%%%%%%%%%%%%%%%%%%

Velocity evolution of runaway bodies in the Case 1 is similar to the 
situation with the intermediate mass spectrum. One finds that
when mass $x$ satisfies $x_c\lesssim x\lesssim 
x_h\equiv s_0^3x_c^{-3\beta}$ velocity profile is shaped by 
the dispersion-dominated interaction with {\it all} small
mass planetesimals ($x\lesssim x_c$):
\begin{eqnarray}
s(x,\tau)\approx s_z(x,\tau)\approx \frac{s_0(\tau)}{x^{1/2}}
\left(x_c^{1-2\beta}\frac{\tilde M_{2+2\beta}}
{\tilde M_{1+4\beta}}\right)^{1/2},~~~3<\alpha<4,~~
x_c\lesssim x\lesssim x_h.
\label{eq:vel_tail_spec8}
\end{eqnarray}
The biggest contribution to both stirring and dynamical friction 
comes the most massive planetesimals with masses $\sim x_c$.

Going to heavier bodies, $x_h\lesssim x\lesssim x_{shear}\equiv s_0^3$,
one finds scattering to be in mixed state: some 
planetesimals [those heavier than $x_s(x)\equiv
(s_0/x^{1/3})^{1/\beta}$] interact 
with big bodies in the shear-dominated regime while the other part 
(those lighter than $x_s$) is in the dispersion-dominated regime 
relative to runaway bodies. It turns out that when $3<\alpha<4$
the biggest contributors to both the heating and friction of
big bodies are planetesimals of mass $\sim x_s$. 
Since this is just at the boundary between the shear- and 
dispersion-dominated regimes, one can  
expect $s_z$ to behave in the same way as $s$ does
[similar to (\ref{eq:vel_tail_spec8})]. Indeed, we
find this to be the case (Figure \ref{fig:less4}a):
\begin{eqnarray}
s(x,\tau)\approx s_z(x,\tau)\approx 
\frac{\sqrt{x_s(x)}}{x^{1/6}}\approx 
\frac{s_0^{1/(2\beta)}}{x^{(1+1/\beta)/6}},~~~3<\alpha<4,~~
x_h\lesssim x\lesssim x_{shear}.
\label{eq:vel_tail_spec9}
\end{eqnarray}

Runaway bodies with $x\gtrsim x_{shear}\equiv s_0^3\approx \tau^{3/4}$ 
experience shear-dominated scattering
by smallest planetesimals; using (\ref{eq:velz_spec_tail1.5}) one finds that
\begin{eqnarray}
s(x,\tau)\approx x^{-1/6}\left(\frac{M_2}{M_1}\right)^{1/2},~~~~~~
s_z(x,\tau)\approx s_0(\tau)x^{-1/2},~~~\alpha>4,~~~x\gtrsim x_{shear}.
\label{eq:vel_tail_spec7}
\end{eqnarray}

%%%%%%%%%%%%%%%%%%%%%%%%%%%%%%%%%%%%%%%%%%%%%%%%%%%%%%%%%%%%%%%%%%%

{\bf Case 2 ($\alpha>4$).}

%%%%%%%%%%%%%%%%%%%%%%%%%%%%%%%%%%%%%%%%%%%%%%%%%%%%%%%%%%%%%%%%%%%

We saw in \S C.1 that in the 
Case 2 planetesimal 
velocity evolution is determined purely by the smallest bodies. 
As a result, it does not matter whether one studies
the dynamics of planetesimals or runaway bodies --- all the 
conclusions of \S C.1 stay unchanged
as long as the interaction with {\it smallest} planetesimals occurs in 
the dispersion-dominated regime. Thus we predict that for 
$x\lesssim x_{shear}$ velocity spectrum is still given by
(\ref{eq:vel_spec5}).  
Runaway bodies of larger mass, $x\gtrsim x_{shear}$, feel
shear-dominated scattering by {\it all} 
planetesimals and their velocity dispersions are given by 
(\ref{eq:vel_tail_spec7}) (exactly like in Case 1).

\begin{center}
\begin{deluxetable}{ l l l l l }
%\tablecolumns{5}
\tablewidth{0pc}
\tablecaption{Summary of analytical results for the velocity 
scaling with mass. 
\label{table2}}
\tablehead{
\colhead{Range of $\alpha$}&
\colhead{Mass interval}&
\colhead{$\eta=d\ln s/d\ln x$}&
\colhead{$\eta_z=d\ln s_z/d\ln x$}&
\colhead{Reference}
}
\startdata
$0<\alpha<2$  & $x\lesssim x_c$ & 
$0$ & $0$ & Eq. (\ref{eq:vel_spec1}) \\
 & $x_c\lesssim x\lesssim x_{shear}$ & 
$-1/2$ & $-1/2$ & Eq. (\ref{eq:vel_spec_tail1})  \\
 & $x\gtrsim x_{shear}$ & 
$-1/6$ & $-1/2$  & Eq. (\ref{eq:velz_spec_tail2}) \\
%%%%%%%%%%%%%%%%%%%%
$2<\alpha<3$  & $x\lesssim x_k$ & 
$0$ & $0$ & Eq. (\ref{eq:vel_spec2})  \\
 & $x_k\lesssim x\lesssim x_c$ & 
$-1/4$ & $-1/4$ & Eq. (\ref{eq:vel_spec3})  \\
 & $x_c\lesssim x\lesssim x_h$ & 
$-1/2$ & $-1/2$ & Eq. (\ref{eq:vel_spec_tail3})  \\
 & $x_h\lesssim x\lesssim x_{shear}$ & 
$-5/6 $ & $-7/6~~~~~~~~~~(\alpha<5/2)$ 
& Eq. (\ref{eq:vel_spec_tail4}) \& 
(\ref{eq:vel_spec_tailz4.5})\\
 &  & $-5/6 $ & $(4\alpha-17)/6~~(\alpha>5/2)$ 
& Eq. (\ref{eq:vel_spec_tail4}) \& 
(\ref{eq:vel_spec_tailz5})  \\
 & $x\gtrsim x_{shear}$ & 
$-1/6$ & $-1/2$ & Eq. 
(\ref{eq:vel_spec_tail6})-(\ref{eq:vel_spec_tailz7})  \\
%%%%%%%%%%%%%%%%%%%%
$3<\alpha<4$  & $x\lesssim x_c$ & 
$-\beta$ & $-\beta$ & Eq. (\ref{eq:I1I2_cond})  \\
 & $x_c\lesssim x\lesssim x_h$ & 
$-1/2$ & $-1/2$ & Eq. (\ref{eq:vel_tail_spec8})  \\
 & $x_h\lesssim x\lesssim x_{shear}$ & 
$-(1+1/\beta)/6$ & $-(1+1/\beta)/6$ & Eq. 
(\ref{eq:vel_tail_spec9})  \\
 & $x\gtrsim x_{shear}$ & 
$-1/6$ & $-1/2$ & Eq. (\ref{eq:vel_tail_spec7})  \\
%%%%%%%%%%%%%%%%%%%%
$\alpha>4$  & $x\lesssim x_{shear}$ & 
$-1/2$ & $-1/2$ & Eq. (\ref{eq:vel_spec5}), 
\S C.2 \\
 & $x\gtrsim x_{shear}$ & 
$-1/6$ & $-1/2$ & \S C.2 \\
\enddata
\end{deluxetable}
\end{center}

\end{document}